\documentclass[floatfix,%
 reprint,
nofootinbib,
 amsmath,amssymb,
 aps,
floatfix,
]{revtex4-2}

\usepackage{graphicx}
\usepackage{braket}
\usepackage{subcaption}
\usepackage{xcolor}
    \usepackage{framed}% Include figure files
\usepackage{dcolumn}% Align table columns on decimal point
\usepackage{bm}
\usepackage{float}
\usepackage{microtype}% bold math
\usepackage{mathrsfs}
\usepackage[colorlinks,citecolor=blue,urlcolor=blue,linkcolor=blue]{hyperref}% add hypertext capabilities

\begin{document}

\preprint{APS/123-QED}

\title{Gravitational Collapse and Singularity Formation in Brans-Dicke Gravity}% Force line breaks with \\
%\thanks{Our work is majorly motivated by the work done in \cite{Goswami:2005fu, goswami2004nakedsingularityformationscalar} and \cite{mosani2023gravitationalcollapsescalarvector}.}%

\author
{Ayush Bidlan$^{1}$}
\email{i21ph018@phy.svnit.ac.in}
\author{ Dipanjan Dey$^{2}$}%
\email{deydipanjan7@gmail.com}
\author{Parth Bambhaniya$^{3}$}
\email{grcollapse@gmail.com}
\affiliation{%
 $^{1}$Department of Physics, Sardar Vallabhbhai National Institute of Technology, Surat 395007, Gujarat, India,
}%
\affiliation{$^{2}$Beijing Institute of Mathematical Sciences and Applications, Beijing 101408, China,}
\affiliation{$^{3}$Institute of Astronomy, Geophysics and Atmospheric Sciences, University of São Paulo, IAG, Rua do Matão 1225, CEP: 05508-090 São Paulo - SP - Brazil}%%

\date{\today}

\begin{abstract}
We investigate gravitational collapse driven solely by a self-interacting Brans–Dicke (BD) scalar field in the absence of ordinary matter. In this framework, the spacetime dynamics are governed solely by the scalar field $\Phi$, endowed with a self-interaction potential $V(\Phi)$ and non-minimally coupled to the Ricci scalar through the Brans–Dicke action. We numerically solve for the evolution of $\Phi(t)$ and the corresponding potential $V(\Phi)$ in order to track the collapse dynamics leading to singularity formation. Our analysis demonstrates that, for the energy densities  $\rho \approx 1/a$ and $\rho \approx -\ln a$, the collapse inevitably leads to the formation of a central curvature singularity while consistently satisfying the weak energy condition. We further examine the causal structure of the resulting singularity and find that future-directed null geodesics originating from the singularity can propagate to future null infinity, making the singularity globally visible. The strength of the singularity is also examined by extending Tipler’s strong curvature condition to the Brans–Dicke field equations. Overall, our findings indicate that gravitational collapse in scalar–tensor gravity can give rise to scenarios that challenge the Cosmic Censorship Conjecture, while underscoring the potential observational relevance of singularities formed through BD scalar-field–driven collapse.

\end{abstract}


\maketitle

\section{Introduction}

Gravitational collapse is one of the most fundamental processes in relativistic astrophysics, forming the basis for our understanding of compact object formation, singularities, and the limits of Einstein's general relativity (GR). In simple terms, continual gravitational collapse refers to the persistent contraction of matter under its own gravity, eventually leading to regions of very high spacetime curvature. The fundamental work of Oppenheimer, Snyder, and Datt (OSD) \cite{Oppenhm} provided the first exact relativistic model for this process. In their model, a spherically symmetric dust cloud of homogeneous density undergoes continual contraction until it reaches a singular state. During the OSD collapse, trapped surfaces appear, which causally disconnect the central region from the outside universe. This ensures that the final singularity is hidden behind an event horizon, resulting in the formation of a black hole (BH). An \textit{event horizon} is defined as the boundary of the causal past of future null infinity ($\mathscr{I}^+$). Therefore, any event located inside the event horizon cannot be causally connected to distant observers outside the event horizon, making the horizon a one-way causal boundary that hides singularities or other strong field regions from the external universe. In this framework, the singularity represents the breakdown of the spacetime manifold itself, where non-spacelike geodesics terminate, and physical quantities such as energy density and curvature become infinite.

While the OSD model is foundational, it is highly idealized. The assumptions of perfect homogeneity and vanishing pressure are rarely realized in actual astrophysical systems. To overcome these limitations, subsequent studies extended collapse models to incorporate inhomogeneous matter distributions, nonzero pressure, and more general equations of state. At a more fundamental level, the singularity theorems of Penrose and Hawking \cite{hawking1970singularities,Hawking:1972qk,penrose2019singularity} demonstrated that singularities are unavoidable under generic conditions. However, these theorems do not determine whether the resulting singularities are visible. Penrose later proposed the Cosmic Censorship Conjecture (CCC) \cite{penrose}, which posits that singularities formed in gravitational collapse should be causally disconnected from distant observers. Despite many decades of research, the CCC has neither been rigorously proved nor disproved, and remains one of the central open problems in gravitational collapse theory.

Considerable attention has been given to gravitational collapse scenarios in GR that extend beyond the idealised OSD model. These studies have shown that, depending on the initial conditions of the collapsing matter, certain dynamics of trapped surfaces inside the collapsing matter allow the central singularity to be visible at least locally \cite{Joshi:1993zg,Joshi:2011zm,Joshi:2011rlc,Joshi:2024gog,Joshi:2013dva, mosani2022aspectsvisiblesingularitiesgravitational,Mosani_2022,mosani, Dey_2022,Mosani:2021byw,Mosani:2020ena}.
The possible existence of naked singularities challenges the CCC and raises questions about the completeness of GR as a theory of gravity. The implications of naked singularities are significant because if they exist, they could produce observable effects that differ from black holes. The proposed observational signatures include changes in the properties of the shadow and accretion disk \cite{Saurabh:2023otl,Solanki:2021mkt,Bambhaniya:2021ybs,Vagnozzi:2022moj}, deviations in the periastron precession of relativistic orbits \cite{Bambhaniya:2019pbr,Bambhaniya:2020zno,Bambhaniya:2025xmu,Bambhaniya:2022xbz,Bambhaniya:2021jum}, pulsar timing variations \cite{Bambhaniya:2025qoe,Kalsariya:2024qyp}, and effects of the tidal force \cite{Madan:2022spd}. Collectively, these findings suggest that naked singularities, if formed in nature, could act as strong field laboratories that provide insights that are not accessible in standard black hole physics \cite{Acharya:2023vlv}.

Although GR has successfully passed a wide range of experimental and observational tests, it is not free from conceptual and phenomenological limitations. Outstanding puzzles such as the origin of dark energy and dark matter, as well as the incompatibility of GR with quantum field theory, have long motivated the exploration of alternative formulations of gravity \cite{Faraoni:2004pi}. Among these, the BD scalar-tensor theory \cite{Brans:1961sx} has played a central role since its inception, originally conceived as a relativistic implementation of Mach’s principle \cite{Godel:1949ga}. In BD gravity, the Newtonian constant $G$ is promoted to a dynamical scalar field $\Phi$, non-minimally coupled to the spacetime curvature via the dimensionless coupling parameter $\omega$. Gravitational interactions are therefore jointly mediated by the metric and the scalar sector, introducing qualitatively new features absent in Einstein gravity. 

Over the decades, BD theory has been extensively applied across cosmology and astrophysics. It has been employed to model the late-time acceleration of the Universe \cite{Elizalde_2004,Raveri_2017}, drive inflationary dynamics in the early Universe \cite{Barrow:1995fj,Garc_a_Bellido_1995}, and examine gravitational collapse and singularity formation \cite{Rasouli_2016,Ziaie:2010cz,Ziaie_2024}. However, solar system and astrophysical observations impose stringent lower bounds on the BD parameter $\omega$ (typically $\omega \gtrsim 10^{4}$), strongly constraining deviations from GR in the weak-field regime \cite{Perivo,Mahm}. These bounds, however, apply only when the scalar field is effectively massless or has a mass below a certain upper limit. Consequently, it is primarily in the strong-field regime that one may hope to find observational evidence for BD theory, if it indeed governs gravity in nature. This regime also provides a natural testing ground for quantum-gravity-inspired extensions \cite{BDQG1,BDQG2,BDQG3}. It is well established that Einstein gravity permits only quadrupole (and higher) modes of gravitational radiation. On the other hand, scalar-tensor models predict additional monopole and dipole contributions \cite{Taylor:1982zz,Scheel_1995}. Such deviations could leave distinctive imprints on gravitational-wave observations, accretion physics, and compact object phenomenology, rendering BD theory an especially powerful framework for probing new physics.

In this work, we investigate a particularly distinctive collapse scenario in BD gravity: gravitational collapse driven solely by the BD scalar field, with no contribution from ordinary matter. By setting the matter Lagrangian $\mathcal{L}_{\text{matter}} \to 0$, the stress-energy tensor of the scalar field, $T^{\Phi}_{\mu\nu}$, becomes the sole source of spacetime curvature. The present work focuses on purely scalar-field-driven collapse in BD gravity, examining the dynamical formation of singularities and their causal structure, including the potential emergence of naked singularities, which cannot occur in standard spherically symmetric vacuum spacetimes in GR. Our objective is to determine whether such scalar-field collapse can end in curvature singularities and, if so, whether these singularities are hidden behind horizons (black holes) or remain visible (naked singularities). BD gravity introduces an additional degree of freedom through the scalar field, which can influence strong-field phenomena such as gravitational waves, compact object formation, and near-horizon dynamics. Studying collapse driven solely by the BD scalar field therefore provides a valuable test of scalar-tensor gravity in the strong-field regime. Although direct observational signatures of such scenarios are not yet established, purely scalar-field collapse may, in principle, produce phenomenological features distinct from standard relativistic collapse in GR, motivating future efforts to identify measurable differences.

We have the following arrangement of the paper. In section \ref{2}, we provide a brief review of BD gravity and derive the effective dynamical equations for the scalar field in the absence of matter. Section \ref{III} presents our numerical study of continuous collapse, including the behaviour of curvature invariants. In section \ref{IV}, we analyse the causal structure of the resulting singularities by studying non-spacelike geodesics. Section \ref{V} discusses the junction conditions required to connect the collapsing interior to an exterior Schwarzschild spacetime. In section \ref{VI}, we investigate the strength of the singularity in the sense of Tipler and Kr\'{o}lak, generalised to BD gravity. Finally, in section \ref{VII}, we discuss our results and conclusions. Throughout, we adopt the metric signature $(-,+,+,+)$ and set $c=1$.

\section{Brans-Dicke Theory}\label{2}
The action of the Brans-Dicke theory, with self-interaction potential $V(\Phi)$ and with the matter fields, is given as
\begin{equation}\label{action}
\begin{split}
        \mathcal{S}=\frac{1}{2}\int d^{4}x\sqrt{-g}\left(\Phi R-\frac{\omega}{\Phi}g^{\mu\nu}\nabla_{\mu}\Phi\nabla_{\nu}\Phi-V(\Phi)\right)\\+\int d^{4}x\sqrt{-g}\hspace{1mm}\mathcal{L}_{\text{matter}}
\end{split}
\end{equation}
where $\Phi$ is the BD scalar field, and $\omega$ is the dimensionless BD parameter controlling the coupling strength between the scalar field and spacetime curvature, which is also known as the BD parameter. Here, $\mathcal{L}_{\text{matter}}$ is the Lagrangian of ordinary matter (any form of matter other than the BD scalar field $\Phi$), and $\Phi$ is related to Newton's gravitational constant as $G = 1/\Phi$. In this work, we consider the vacuum case ($\mathcal{L}_{\text{matter}} = 0$) and show that black holes or naked singularities can form due to the coupling between the BD scalar field and curvature. The field equations from this action (using $\delta\mathcal{S}/\delta g_{\mu\nu}=0$) are given as 
\begin{equation}\label{FieldEqn1}
\begin{split}
    R_{\mu\nu}-\frac{1}{2}g_{\mu\nu}R=\frac{\omega}{\Phi^{2}}\left(\nabla_{\mu}\Phi\nabla_{\nu}\Phi-\frac{1}{2}g_{\mu\nu}\nabla_{\eta}\Phi\nabla^{\eta}\Phi\right)+\\\frac{1}{\Phi}\left(\nabla_{\mu}\nabla_{\nu}\Phi-g_{\mu\nu}\Box\Phi\right)-g_{\mu\nu}\frac{V(\Phi)}{2\Phi}.
\end{split}
\end{equation}
Moreover, on applying the variational action principle with respect to the field $\Phi$, i.e. $\delta\mathcal{S}/\delta\Phi=0$, we obtain the massive Klein-Gordon equation as
\begin{equation}\label{FieldEqn2}
    \ddot{\Phi}+3H\dot{\Phi}=-\frac{1}{2\omega+3}\left(\Phi\frac{dV(\Phi)}{d\Phi}-2V(\Phi)\right).
\end{equation}
One then has the following equations for the energy density and pressure as a function of the BD scalar field from its $T^{\Phi}_{\hspace{1mm}\mu\nu}$ stress-energy tensor components 
\begin{equation}\label{BD-Density}
    \rho=\frac{\omega}{2\Phi^{2}}\dot{\Phi}^{2}-3H\frac{\dot{\Phi}}{\Phi}+\frac{V(\Phi)}{2\Phi},
\end{equation}
\noindent and,
\begin{equation}\label{BD-Pressure}
    p=\frac{\ddot{\Phi}}{\Phi}+\frac{\omega}{2\Phi^{2}}\dot{\Phi}^{2}+2H\frac{\dot{\Phi}}{\Phi}-\frac{V(\Phi)}{2\Phi}.
\end{equation}
Here, $H=\dot{a}/a$ is the Hubble parameter. All other off-diagonal terms of the stress-energy-momentum tensor are zero due to homogeneity and isotropy conditions. Let us now consider the interior spacetime of a collapsing scenario, described by the spatially flat Friedmann-Robertson-Walker (FRW) metric,
\begin{equation}\label{FRW}
    ds^{2}=-dt^{2}+a^{2}(t)\left(dr^{2}+r^{2}d\Omega^{2}\right),
\end{equation}
is a solution of the BD field equations Eq.~(\ref{FieldEqn1}).
Throughout, we assume the weak energy condition holds for any timelike observer,
\begin{equation}\label{wec}
\rho\geq0;\hspace{1mm}\rho+p\geq0,
\end{equation}
where
%\textcolor{red}{We then proceed to calculate the field equations for BD gravity using the flat FRW metric in Eq. (\ref{FRW}) in terms of the scale factor $a(t)$ and the scalar field $\Phi(t)$ directly, instead of the LTB-type mass function, as}
\begin{equation}\label{rho}
\rho=\frac{\mathcal{M'}}{\mathcal{R}^{2}\mathcal{R}'},
\end{equation}
\begin{equation}\label{presure}
    p=-\frac{\dot{\mathcal{M}}}{\mathcal{R}^{2}\dot{\mathcal{R}}},
\end{equation}
and,
\begin{equation}\label{FieldEqn3}
    \dot{\mathcal{R}}^{2}=\frac{\mathcal{M}}{\mathcal{R}}.
\end{equation}
The quantity $\mathcal{M}$ represents the Misner-Sharp mass term with $\mathcal{M} \geq 0$, and $\mathcal{R}(t,r)=r~a(t)$ is the area radius for the shell labelled by the comoving coordinate $r$, where at the initial surface we assume $\mathcal{R}(0,r)=r$, i.e., $a(t=0)=1$. By algebraically solving Eq. (\ref{FieldEqn1}), we obtain at a relation between $\dot{a}$ and the  energy density parameter as
\begin{equation}\label{adot}
    \dot{a}=-a\sqrt{\frac{\rho}{3}}.
\end{equation}
Here, the negative sign in Eq. (\ref{adot}) implies that the area radius of the shell for a fixed $r$ decreases monotonically due to the continual collapse of the BD scalar field. In the present homogeneous scenario, a shell of given comoving radius $r$, becomes singular at time $t_s$ when $a(t_s)=0$. Since in homogeneous collapse, the scale factor $a(t)$ is independent of the comoving radius $r$, all the shells inside the spherically symmetric collapsing matter become singular simultaneously. At these types of shell-focusing singularities, curvature and all the physical parameters become unbounded and infinite. Now, one can also obtain the second derivative of the scale factor by differentiating Eq. (\ref{adot}) with respect to time and using the chain rule,
\begin{equation}\label{adotdot}
    \ddot{a}=a \left(\frac{\rho}{3}-\frac{\dot{a}}{2 \sqrt{3\rho}}\frac{d\rho}{da}\right).
\end{equation}
Having explicit expressions for $\dot{a}$ and $\ddot{a}$ allows us to analyze the dynamical equations and investigate whether their solutions indicate the formation of a singularity. In the next section, we examine the collapsing dynamics that can lead to a spacetime singularity in Brans–Dicke gravity, assuming a specific dependence of the energy density $\rho$ on the scale factor $a(t)$ in the vicinity of the singularity (see \cite{goswami2004nakedsingularityformationscalar} for an analogous construction in general relativity). 
\begin{figure*}[t]
    \centering
    \begin{minipage}[b]{0.47\textwidth}
        \centering
        \includegraphics[width=\textwidth]{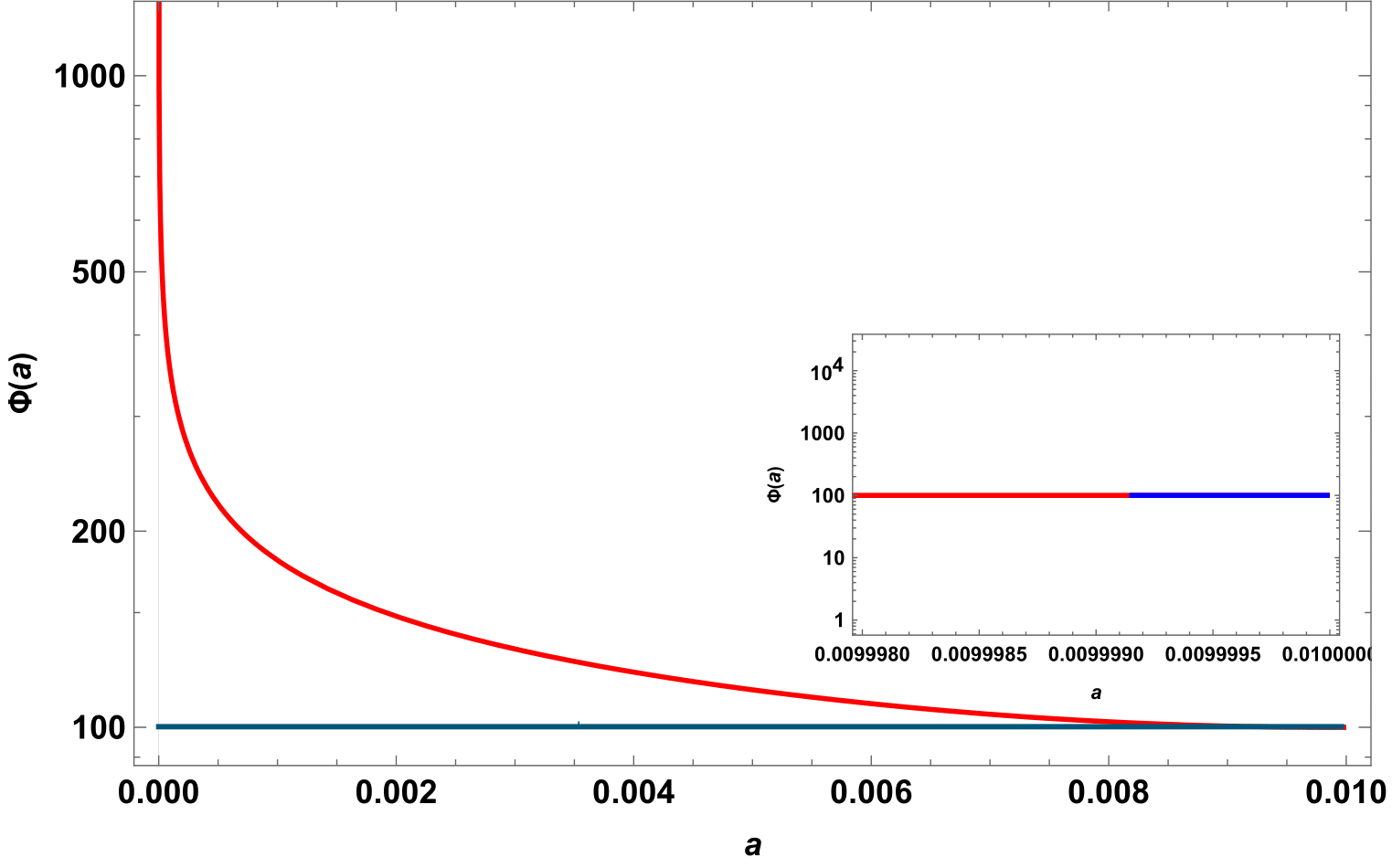}
        \caption{Variation of $\Phi(a)$ with $a(t)$ for $\rho=1/a$ when $\omega=6$ (red) and $\omega=10^{6}$ (blue).}
        \label{Figure 1}
    \end{minipage}
    \hfill
    \begin{minipage}[b]{0.47\textwidth}
        \centering
        \includegraphics[width=\textwidth]{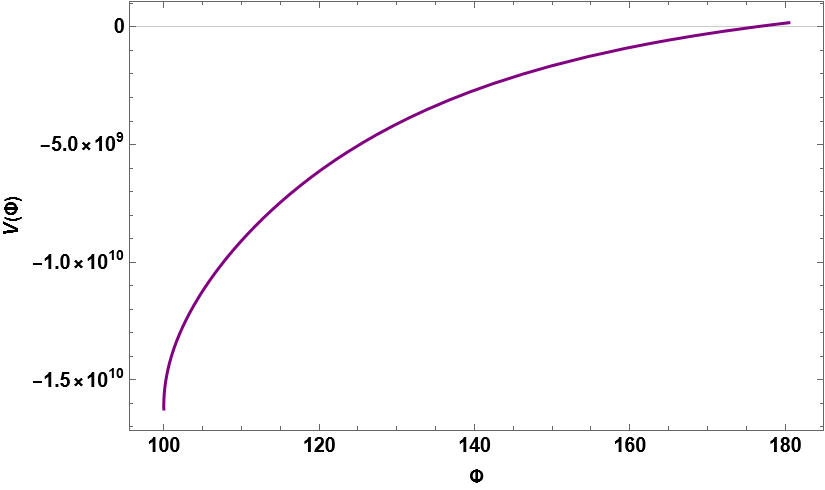}
        \caption{Plot for the potential $V(\Phi)$ against the BD scalar field $\Phi$ for $\rho=1/a$ when $\omega=6$.}
        \label{Fig2}
    \end{minipage}
\end{figure*}
\section{Curvature Singularity from Brans–Dicke Collapse}\label{III}
Before proceeding further, we emphasize that our objective is to formulate the differential equations in a manner that allows for a numerical investigation of general forms of energy density and pressure. At a later stage, these general functions will be approximated by specific expressions. In particular, we shall prescribe a suitable form of $\rho$ in the vicinity of the singularity and subsequently determine $p$ using Eq.~(\ref{FieldEqn2}). Therefore, we proceed by modifying Eqs. (\ref{BD-Density}) and (\ref{BD-Pressure}), and expressing the energy density and pressure as a function of the scale factor $a(t)$ in the following manner
\begin{equation}\label{BD-Density(a)}
    \rho=\frac{\omega\dot{a}^{2}}{2\Phi^{2}}\left(\frac{d\Phi}{da}\right)^{2}-3\frac{\dot{a}^{2}}{a\Phi}\frac{d\Phi}{da}+\frac{V(\Phi)}{2\Phi},
\end{equation}
and,
\begin{equation}\label{BD-Pressue(a)}
    p=\frac{\dot{a}^{2}}{\Phi}\frac{d^{2}\Phi}{da^{2}}+\frac{\omega\dot{a}^{2}}{2\Phi^{2}}\left(\frac{d\Phi}{da}\right)^{2}+\frac{\ddot{a}}{\Phi}\frac{d\Phi}{da}+2\frac{\dot{a}^{2}}{a\Phi}\frac{d\Phi}{da}-\frac{V(\Phi)}{2\Phi}.
\end{equation}
We now write the Misner-Sharp mass function $\mathcal{M}$ as a function of the scale factor $a(t)$ upon solving Eq. (\ref{rho}). Integrating Eq. (\ref{rho}) on both sides, we obtain
\begin{equation}\label{Mass Function}
\begin{split}
    \mathcal{M}=\frac{\mathcal{R}^{3}}{3}\rho.
    \end{split}
\end{equation}
On solving Eqs. (\ref{FieldEqn2}) and (\ref{Mass Function}) together, we obtain the explicit dependence of the pressure $p$ on scale factor $a(t)$ as
\begin{equation}\label{Pressure(singular)}   p=-\left(\rho+\frac{a}{3}\frac{d\rho}{da}\right).
\end{equation}
Consequently, one can obtain the expression for the Equation of State (EoS), namely, $w$, in this scenario as
\begin{equation}\label{EoS}
    w=-1-\frac{a}{3\rho}\frac{d\rho}{da}.
\end{equation}
For the weak energy condition to hold for any local timelike observer, the following inequality must be satisfied, as obtained after using Eq. (\ref{Pressure(singular)}) and Eq. (\ref{wec}) together, as
\begin{equation}
    \frac{d\rho}{da}\leq0.
\end{equation}
As we proceed, we will assume explicit forms for $\rho$, motivated by the fact that the energy density diverges as the collapse advances, while also commenting on the validity of the weak energy condition whenever relevant. Now, we want to study the cosmological evolution of the BD scalar field $\Phi$ during the collapse and in order to retrieve that we add the dynamical equations, i.e., Eqs. (\ref{BD-Density(a)}) and Eq. (\ref{BD-Pressue(a)}), along with using Eqs. (\ref{adot}), (\ref{adotdot}) and (\ref{Pressure(singular)}). We obtain the following non-linear second-order differential equation,
\begin{equation}\label{de3}
   \begin{split}
    \frac{a^{2}\rho}{3\Phi}\frac{d^2\Phi}{da^2}+\frac{\omega\rho a^{2}}{3\Phi^{2}}\left(\frac{d\Phi}{da}\right)^{2}+\frac{a^{2}\rho_{,a}}{6\Phi}\frac{d\Phi}{da}+\frac{a}{3}\frac{d\rho}{da}=0.
   \end{split}
\end{equation}

The above differential equation is highly non-linear, and obtaining an analytic solution is non-trivial. Therefore, we consider specific forms of the energy density $\rho$ near the central singularity and subsequently solve the above differential equation numerically in the limit $a\to 0$, and plot the corresponding evolution of $\Phi$ as a function of $a$ in this limit. As a starting point, we adopt a simple choice, also considered in \cite{goswami2004nakedsingularityformationscalar}, namely $\rho \sim 1/a$ as $a\to 0$. At a later stage, we will also consider an alternative form, where the energy density is taken to be a negative logarithmic function of the scale factor, i.e., $\rho = -\ln a$ in the same limit of $a$ mentioned above. Let us, for now, consider an ansatz for the energy density as follows
\begin{equation}
    \rho=\frac{1}{a}.
\end{equation}
By using Eq. (\ref{Pressure(singular)}), the corresponding expression for the pressure becomes
\begin{equation}
    p=-\frac{2}{3a}.
\end{equation}
Moreover, the EoS parameter for this scenario becomes
\begin{equation}
    w=-\frac{2}{3}.
\end{equation}

This suggests that the weak energy condition is satisfied, i.e., $w\geq-1$ for all values of the scale factor $a$ with this particular form of the expression for energy density. Furthermore, with $\rho=1/a$, the corresponding form of the differential equation (\ref{de3}) becomes the following:
\begin{equation}\label{de4}
   \begin{split}
    \frac{a}{3\Phi}\frac{d^2\Phi}{da^2}+\frac{\omega a}{3\Phi^{2}}\left(\frac{d\Phi}{da}\right)^{2}-\frac{1}{6\Phi}\frac{d\Phi}{da}-\frac{1}{3a}=0.
   \end{split}
\end{equation}
  Figure ~(\ref{Figure 1}) illustrates the dynamics of the BD scalar field $\Phi(a)$ as a function of the scale factor $a$, for the coupling parameter $\omega = 6$ and $\omega = 10^{6}$.  The latter value is chosen according to the observational constraint on $\omega$ discussed in the Introduction.
 The evolution of the BD scalar field shows a highly divergent nature as $a\to 0$, suggesting a significant change in the gravitational force due to the inverse relation between the BD scalar field and the gravitational constant.
 
\begin{figure*}[t]
    \centering
    \begin{minipage}[b]{0.47\textwidth}
        \centering
        \includegraphics[width=\textwidth]{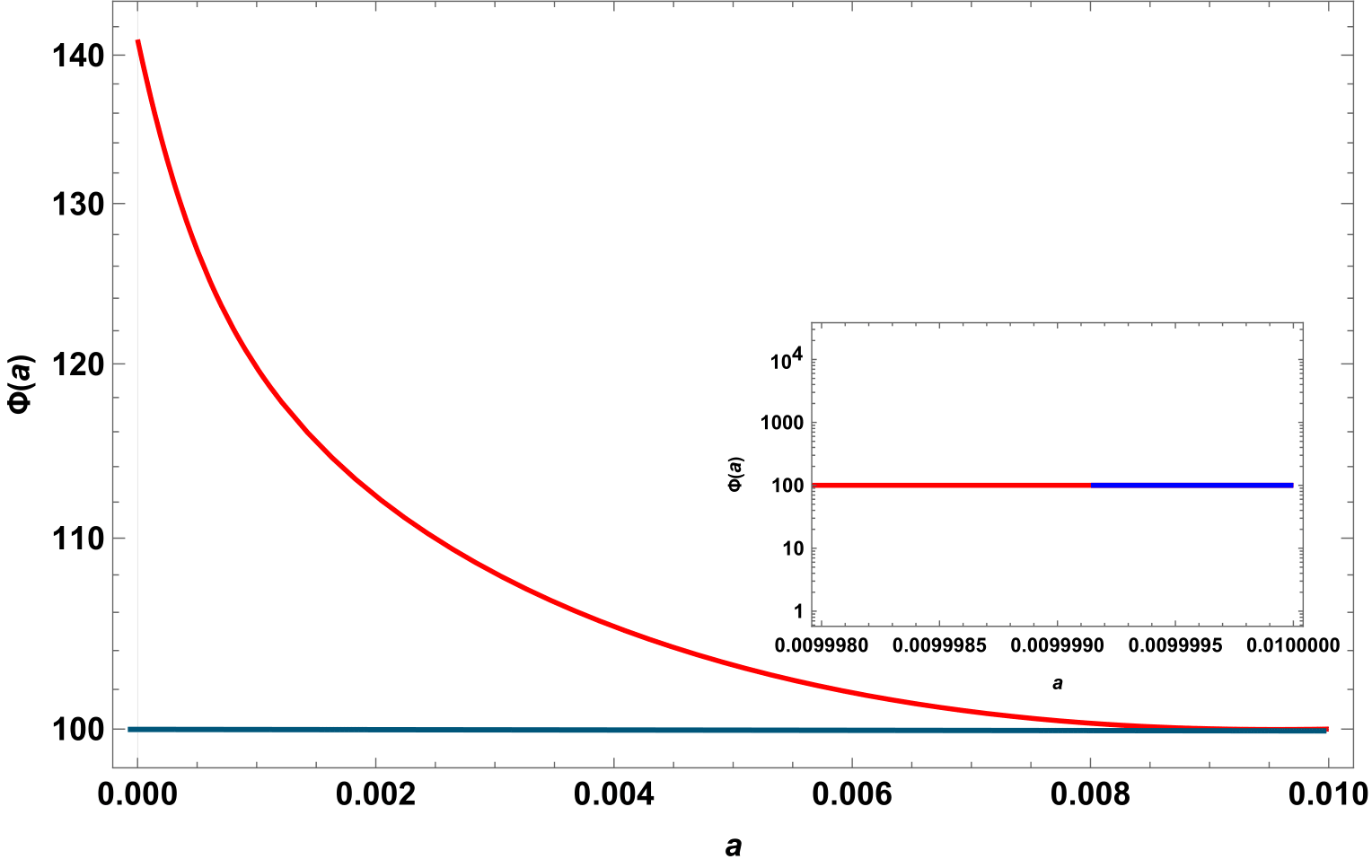}
        \caption{Variation of $\Phi(a)$ with $a(t)$ for $\rho=-\ln{a}$ when $\omega=6$ (red) and $\omega=10^{6}$ (blue).}
        \label{Fig3}
    \end{minipage}
    \hfill
    \begin{minipage}[b]{0.47\textwidth}
        \centering
        \includegraphics[width=\textwidth]{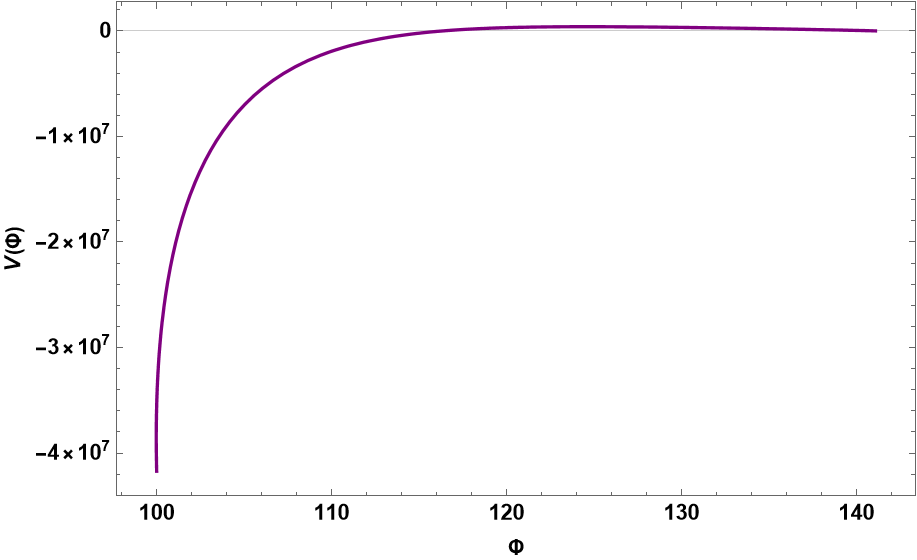}
        \caption{Plot for $V(\Phi)$ against the BD scalar field $\Phi$ for $\rho=-\ln{a}$ when $\omega=6$.}
        \label{Fig4}
    \end{minipage}
\end{figure*}
Our next objective is to obtain the numerical solution for the potential \(V(\Phi)\), which characterizes the self-interaction of the Brans–Dicke (BD) scalar field. Subtracting Eq.~(\ref{BD-Density(a)}) from Eq.~(\ref{BD-Pressue(a)}) yields  
\begin{equation}
\begin{split}
    V(\Phi) = \frac{a^{2}\rho}{3}\frac{d^{2}\Phi}{da^{2}} 
    + \left( 2a\rho + \frac{a^{2}\rho_{,a}}{6} \right)\frac{d\Phi}{da} 
    + 2\rho\Phi + \frac{a\Phi}{3}\frac{d\rho}{da},
\end{split}
\label{Vphigen}
\end{equation}
which, for the choice \(\rho = 1/a\), reduces to  
\begin{equation}
    V(\Phi) = \frac{a}{3}\frac{d^{2}\Phi}{da^{2}} 
    + \frac{11}{6}\frac{d\Phi}{da} 
    + \frac{5\Phi}{3a}.
\end{equation}
Using the above expression together with the numerical solution of $\Phi(a)$ obtained from Eq.~(\ref{de4}), we determine the corresponding numerical behavior of $V(\Phi)$. Figure~(\ref{Fig2}) illustrates the variation of $V(\Phi)$ with respect to $\Phi$ obtained from this numerical analysis. It can be observed that the potential $V(\Phi)$ increases monotonically, remaining negative up to a certain value of $\Phi$, beyond which it becomes positive as $\Phi$ continues to increase.
 Since $\Phi(a)$ increases during collapse (i.e., as $a$ decreases; see Fig.~(\ref{Figure 1})), $V(\Phi)$ also increases with the progression of collapse. For $\omega = 10^{6}$, the Brans–Dicke scalar field $\Phi$ exhibits negligible variation with respect to the scale factor $a$, and consequently the potential $V(\Phi)$ remains nearly constant throughout the evolution. Therefore, the corresponding potential is not shown in Fig.~(\ref{Fig2}).

As mentioned before, another functional form of the energy density that we consider is
$\rho = -\ln{a}$.
In this case, the equation-of-state (EoS) parameter becomes
$$w = -1 - \frac{1}{3\ln{a}}.$$
It can be seen from the above expression of $w$ that the weak energy condition (WEC) is satisfied for $0\leq a < 1$. Therefore, as in the case of $\rho = 1/a$, it is physically consistent to restrict the analysis to the limit $a \rightarrow 0$, corresponding to the final stage of collapse. Now,
substituting $\rho = -\ln{a}$ into Eq.~(\ref{de3}), we obtain the governing differential equation for the BD scalar field $\Phi(a)$ as
\begin{equation}
    \begin{split}
        \frac{a^2 \ln{a}}{3 \Phi}\frac{d^{2}\Phi}{da^{2}}+\frac{\omega a^2 \ln{a}}{3\Phi^2}\left(\frac{d\Phi}{da}\right)^{2}+\frac{a}{6\Phi}\frac{d\Phi}{da}+\frac{1}{3}=0.
    \end{split}
\end{equation}
Figure~(\ref{Fig3}) shows the numerical evolution of the BD scalar field $\Phi(a)$ for $\rho = -\ln{a}$. The field increases monotonically with decreasing $a$, similar to the $\rho = 1/a$ case (see Fig.~(\ref{Figure 1})), implying that the effective gravitational strength decreases progressively as the collapse advances.
The corresponding potential $V(\Phi)$, obtained from Eq.~(\ref{Vphigen}), is given by
\begin{equation}
\begin{split}
    V(\Phi)=-\frac{a^{2}\ln{a}}{3}\frac{d^{2}\Phi}{da^{2}}-\left(2a\ln{a}+\frac{a}{6}\right)\frac{d\Phi}{da}\\-2\Phi\ln{a}-\frac{\Phi}{3}.
    \end{split}
\end{equation}
Figure~(\ref{Fig4}) depicts the variation of $V(\Phi)$ with $\Phi$ for the same range of $a$ (i.e., $0<a<<1$). The potential remains negative throughout and increases monotonically toward zero as $\Phi$ increases. Therefore, unlike $\rho = 1/a$ case, it does not get positive value for any large value of $\Phi$.
 
Next, we would also like to analyse the Kretschmann scalar for the FRW line element, which can be written as
\begin{equation}
    \kappa\equiv R^{\mu\nu\rho\sigma}R_{\mu\nu\rho\sigma}=\frac{12}{a^{4}}\left((a\ddot{a})^{2}+(\dot{a})^{4}\right)
\end{equation}
Substituting the expressions for $\ddot{a}$ and $\dot{a}$ in their general form produces the following invariant as a function of $\rho$:
\begin{equation}
        \kappa=12\left[\frac{\rho^{2}}{9}+\left(\frac{\rho}{3}+\frac{a}{6}\frac{d\rho}{da}\right)^{2}\right]
\end{equation}
One can now estimate the Kretschmann invariant for the selected form of $\rho$, i.e, $\rho=1/a$ and $\rho=-\ln{a}$. We proceed as follows:
\begin{itemize}
    \item For $\rho=1/a$, the $\kappa$ becomes,
    \begin{equation}
        \kappa=\frac{5}{3a^2}
    \end{equation}
    
    \item For $\rho=-\ln{a}$, the $\kappa$ becomes,
    \begin{equation}
        \kappa=\frac{1}{3}\left(1 + 4\ln{a} + 8 (\ln{a})^2\right)
    \end{equation}
\end{itemize}
%The combined numerical plot for both cases is represented in Figure (\ref{Fig5}). 
It is observed that $\kappa$ diverges as $a\rightarrow0$ for both $\rho=1/a$ and $\rho=-\ln{a}$ cases. 
%At late times, i.e., $a\rightarrow1$, the invariant value converges to zero signalling the failure of the formation of the event horizon. 
In the next section, we examine the causal structure of the singularity to determine whether it is visible or hidden. Specifically, we investigate the existence of outgoing causal curves originating from the singularity.

%\begin{figure}[h]
%    \centering
    %\includegraphics[width=1\linewidth]{plot7.png}
%    \caption{\textcolor{red}{Plot} for the Kretschmann invariant against the scale factor $a$ in the limiting case as $a\rightarrow 0$. The red and blue plot indicate $\rho=1/a$ and $\rho=-\ln{a}$, respectively.}
%    \label{Fig5}
%\end{figure}

\section{Causal Structure of the Singularity}\label{IV}
A singularity formed by unhindered gravitational collapse is termed naked if there exists a family of outgoing causal curves with their past endpoint at the singularity. These curves may either reach a distant observer or return to the singularity itself, giving rise to globally or locally naked singularities, respectively.
The visibility of the singularity is fundamentally governed by the geometry of trapped surfaces that develop during the collapse. Trapped surfaces are two-dimensional spacelike surfaces where both ingoing and outgoing null congruences converge. The convergence or divergence of the outgoing and ingoing null geodesics is characterized by their expansion scalar $\Theta_l$ and $\Theta_n$, respectively.

Let us consider the congruence of outgoing radial null geodesics with tangent vectors $(\xi^{\tau},\xi^{r},0,0)$, where $\xi^{\tau}=\frac{d\tau}{d\lambda}$ and $\xi^{r}=\frac{dr}{d\lambda}$. The geodesic equations are
\begin{equation}
\frac{d\xi^{r}}{d\lambda}=-2\frac{\dot{a}}{a}\xi^{\tau}\xi^{r},
\end{equation}
and
\begin{equation}
\frac{d\xi^{\tau}}{d\lambda}=-a\dot{a}\left(\xi^{r}\right)^{2}.
\end{equation}
The expansion parameter is given by
\begin{equation}
\Theta_l = \nabla_{i}\xi^{i}.
\end{equation}
After evaluating the relevant Christoffel symbols and simplifying, the final expression for $\Theta$ becomes
\begin{equation}
\Theta_l=\frac{2}{\mathcal{R}}\left(1-\sqrt{\frac{\mathcal{M}}{\mathcal{R}}}\right),
\end{equation}
where, as we stated before, $\mathcal{R}$ and $\mathcal{M}$ denote the radius of the area and the Misner–Sharp mass, respectively.

%The visibility of the singularity is determined by the behaviour of trapped surfaces characterized by the condition $\mathcal{M}=\mathcal{R}$, which defines the apparent horizon. 

The absence of trapped surfaces in the vicinity of the singularity permits a family of outgoing null geodesics to escape from it, making the singularity at least locally visible. A trapped surface is a spacelike two-surface on which the expansions of both null congruences satisfy $\Theta_l \le 0$ and $\Theta_n < 0$. The marginal condition $\Theta_l = 0$ and $\Theta_n < 0$ characterizes a marginally trapped surface (MTS), and the foliation of such MTSs generates a three-dimensional hypersurface known as the apparent horizon (AH). In the present scenario, for $\rho=\frac1a$, we can write
$\frac{\mathcal{M}}{\mathcal{R}}=\frac{\mathcal{R}^{2}}{3}\rho=\frac13r^2a$. Now, one can consider a collapsing scenario where a spherically symmetric vacuum region ($T_{\mu\nu}^{(m)}=0$) with a certain comoving boundary radius $r_b$ collapses due to the presence of a non-zero curvature caused by homogeneous BD scalar field. Since $a\in[0,1]$ throughout the collapse, if $r_b<\sqrt3$, $\Theta_l$ remains positive, implying trapped surfaces would not form, and the singularity formed in this collapse becomes globally visible, allowing future-directed non-spacelike geodesics emerging from it to reach the asymptotic null infinity. For $\rho =-\ln{a}$, if the boundary radius $r_b<\frac{\sqrt3}{\max{\left(a\sqrt{|\ln{a}|}\large\right)}}\sim 4.0385$, then there would be no trapped surfaces in the whole collapse dynamics resulting the null geodesics to escape from the central singularity and to reach the asymptotic null infinity.

Now, when $r_b > \sqrt{3}$ for $\rho = 1/a$ and $r_b > 4.0385$ for $\rho = -\ln a$, we obtain $\mathcal{M}/\mathcal{R} \ge 1$ for a finite time interval within $[r_{MTS}(t), r_b]$, where $r_{MTS}(t)$ denotes the radius of the marginal inner trapped surface at time $t$. This indicates the formation of trapped surfaces whose inner boundary satisfies $\mathcal{M} = \mathcal{R}$.
One observes that $r_{MTS}(t)$ increases as the collapse progresses in both cases (i.e., for $\rho = 1/a$ and for $\rho = -\ln a$), eventually reaching $r_b$. Once the marginal inner trapped surface meets the boundary of the collapsing region, that region becomes free again of trapped surfaces, and the final singularity becomes naked. This scenario may have important physical implications. While the trapped band persists in the interval $[r_{MTS}(t), r_b]$, all the outgoing radial null geodesics originating from the inner region may not escape to future null infinity, resulting in a strongly suppressed observed intensity. Once the trapped band disappears, these previously confined null rays are suddenly able to propagate outward, producing an abrupt increase in the observed intensity. Such a sharp transition reflects a change in the internal dynamical state of the collapsing matter or can reveal otherwise inaccessible aspects of the underlying spacetime geometry.

In this section, we have shown that, using an ansatz-based approach, a naked singularity can form as a result of the gravitational contraction of the Brans–Dicke scalar field $\Phi$. In the following section, we construct a spacetime configuration in which the interior is a spatially flat FLRW geometry sourced solely by the Brans–Dicke scalar field, while the exterior is described by a generalized Vaidya spacetime that may be sourced by both the Brans–Dicke scalar field and ordinary matter.

\section{Junction Conditions in Brans-Dicke Gravity}
\label{V}

In Brans-Dicke (BD) gravity, the matching of two spacetimes across a timelike hypersurface $\Sigma$ is governed by generalized junction conditions that extend the standard Israel-Darmois conditions of general relativity. These conditions ensure a consistent matching of both the spacetime geometry and the BD scalar field across the boundary.

Let $h_{ij}$ be the induced metric on $\Sigma$, $n^\mu$ the unit normal to $\Sigma$, $K_{ij}$ the extrinsic curvature, and $\Phi$ the Brans-Dicke scalar field. The jump of any quantity $X$ across $\Sigma$ is denoted by $\langle X\rangle \equiv X^{+}-X^{-}$.

The general junction conditions in Brans-Dicke theory are given by the following set of equations:
\begin{enumerate}
\item Continuity of the first fundamental form,
$\langle h_{ij}\rangle=0$.
\item Generalized Israel conditions relating the jump in the extrinsic curvature to the surface stress--energy tensor $S^{i}{}_{j}$ on $\Sigma$,
$-\langle K^{i}{}_{j}\rangle+\delta^{i}_{j}\langle K\rangle
=\frac{1}{\Phi}\left(S^{i}{}_{j}-\frac{S}{3+2\omega}\delta^{i}_{j}\right)$,
where $K=K^{i}{}_{i}$, $S=S^{i}{}_{i}$, and $\omega$ is the Brans-Dicke coupling parameter.

\item Continuity of the scalar field across the hypersurface,
$\langle\Phi\rangle=0$.

\item A jump condition for the normal derivative of the scalar field,
$\langle\partial_{n}\Phi\rangle=\frac{S}{3+2\omega}$.
\end{enumerate}

These equations constitute the complete set of junction conditions in Brans-Dicke gravity for a non-null hypersurface.

\subsection{Specialisation to the Present Collapse Model}

We now apply the above junction conditions to the spherically symmetric gravitational collapse model considered in this work. The interior region $r\le r_b$ is described by a spatially flat FLRW spacetime sourced solely by the Brans–Dicke scalar field $\Phi$, with no ordinary matter present, i.e. $T^{(\mathrm{m})}_{\mu\nu}=0$. The exterior region is described by a generalised Vaidya spacetime, which, within Brans–Dicke gravity, should be interpreted as an effective radiating geometry. In this case the null–fluid–like character of the exterior does not arise solely from ordinary matter, but rather from the combined contribution of the Brans–Dicke scalar field and any matter-sector to the effective stress–energy tensor appearing in the field equations. Consequently, the outgoing null radiation in the exterior encodes the dynamical influence of the scalar field during the late stages of collapse. The generalized Vaidya metric can be written in the following form (\cite{Vaidya:1951zz,Vaidya:1953zza,Vaidya:1999zz}),
\begin{equation}\label{exterior}
    ds^{2}_{+}=-\left(1-\frac{2M(r_{v},v)}{r_{v}}\right)dv^{2}-2dvdr_{v}+r^{2}_{v}d\Omega^{2},
\end{equation}
where $v$ is the retarded null coordinate, $r_{v}$ and $M(r_{v},v)$ are the Vaidya radius and mass, respectively. Having specified the exterior geometry, we now employ the junction conditions discussed above to match the interior collapsing spacetime to the generalized Vaidya exterior across the boundary hypersurface $\Sigma$ defined by $r=r_b$.
The boundary hypersurface $\Sigma:r=r_b$ is assumed to be free of any thin shell of matter. Consequently, the surface stress--energy tensor vanishes,
$S^{i}{}_{j}=0$.
Under this assumption, the Brans-Dicke junction conditions simplify to
$\langle K^{i}{}_{j}\rangle=0$ and $\langle\partial_n\Phi\rangle=0$,
together with the continuity conditions
$\langle h_{ij}\rangle=0$ and $\langle\Phi\rangle=0$.

Thus, both the intrinsic and extrinsic geometries of the hypersurface $\Sigma$ are continuous, and the Brans--Dicke scalar field as well as its normal derivative match smoothly across the boundary. This guarantees that the interior FLRW spacetime sourced by $\Phi$ can be consistently matched to the exterior generalized Vaidya geometry without introducing any surface layer.

%Although the interior region is vacuum in the usual sense of ordinary matter fields, the Brans--Dicke scalar field induces an effective energy--momentum tensor that drives the collapse toward a curvature singularity. For the choice $\omega=6$, the resulting singularity is locally naked. The smooth matching of $\Phi$ and $\partial_n\Phi$ across $\Sigma$ ensures that the scalar field contributes nontrivially to the effective stress--energy of the exterior spacetime, allowing the generalized Vaidya region to encode a flux of outgoing null radiation during the late stages of collapse. This radiation can propagate to distant observers, thereby revealing the locally naked nature of the singularity while maintaining a regular junction across the boundary.
The spacetime metric just inside $\Sigma$ is given by
\begin{equation}\label{interior}
    ds^{2}_{-}=-d\tau^{2}+a^{2}(\tau)\left(dr^{2}+r^{2}_{b}d\Omega^{2}\right),
\end{equation}
whereby matching the area radius at the boundary provides the following condition
\begin{equation}
    R(r_{b},\tau)=r_{v}(v).
\end{equation}
One then gets the interior and exterior induced metrics on the hypersurface $\Sigma$ as follows
\begin{equation}
    ds^{2}_{\Sigma_{-}}=-d\tau^{2}+a^{2}(\tau)r^{2}_{b}d\Omega^{2},
\end{equation}
and,
\begin{equation}
    ds^{2}_{\Sigma_{+}}=-\left(1-\frac{2M(r_{v},v)}{r_{v}}+2\frac{dr_{v}}{dv}\right)dv^{2}+r^{2}_{v}d\Omega^{2},
\end{equation}
where matching gives the first fundamental form as
\begin{equation}\label{1st}
    \left(\frac{dv}{d\tau}\right)_{\Sigma}=\left[1-\frac{2M(r_{v},v)}{r_{v}}+2\frac{dr_{v}}{dv}\right]^{-\frac{1}{2}}.
\end{equation}

\begin{figure*}[t]
\centering
\includegraphics[width=\textwidth]{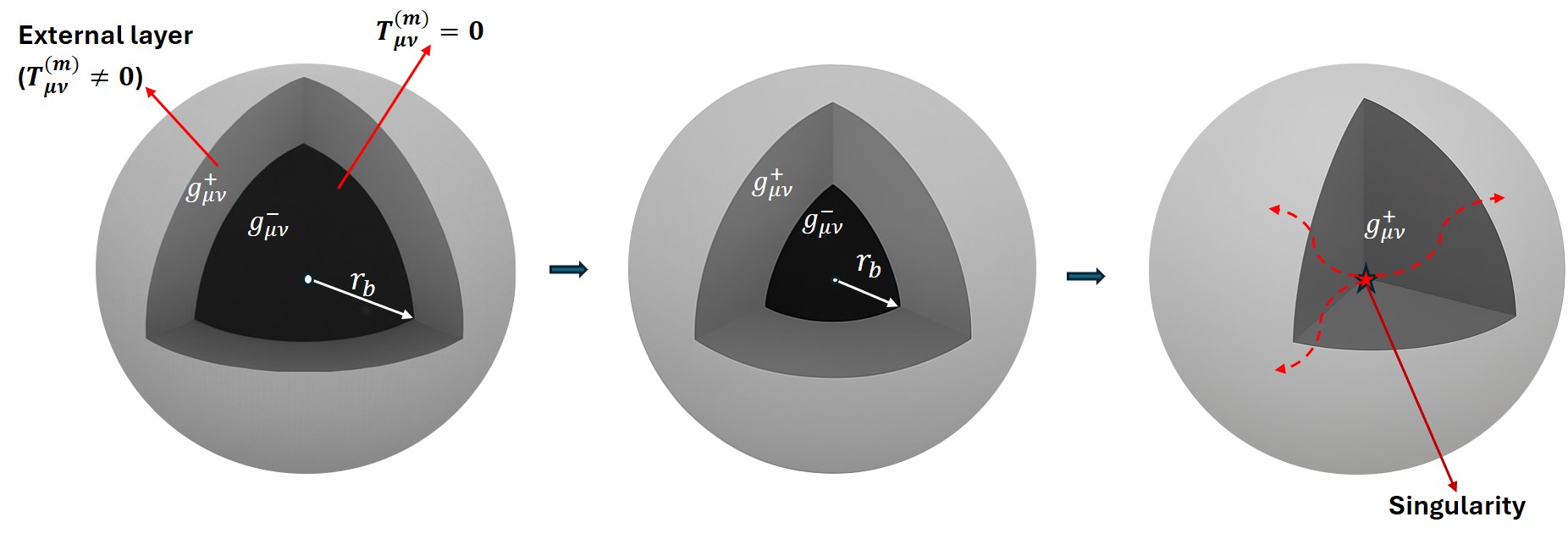}
\caption{A finite interior region ($g^-_{\mu\nu}$), sourced only by the Brans–Dicke scalar field ($T^{(\mathrm m)}_{ab}=0$), is smoothly matched at $r=r_b$ to an exterior generalized Vaidya spacetime ($g^+_{\mu\nu}$). The collapse culminates in a central naked singularity, with outgoing null geodesics escaping to the exterior.}
\label{BDnaked}
\end{figure*}

In order to match the second fundamental form for the extrinsic curvature in the interior and exterior spacetimes, we need to find the unit normal vector field to the hypersurface $\Sigma$. We then proceed by incorporating the fact that any spacetime metric can be written locally in the following form as
\begin{equation}
    ds^{2}=-\left(\alpha^{2}-\beta_{i}\beta^{i}\right)d\tau^{2}-2\beta_{i}dx^{i}d\tau+h_{ij}dx^{i}dx^{j},
\end{equation}
where $\alpha$, $\beta^{i}$ and $h_{ij}$ are the lapse function, shift vector, and induced three-metric, respectively. Here $i,j=1,2,3$. Comparing the interior and the exterior spacetime metrics by using the normalisation condition for $n^{v}$ and $n^{r_{v}}$ as
\begin{equation}
    n_{\mu}n^{\mu}\equiv n^{v}n_{v}+n^{r_{v}}n_{r_{v}}=1,
\end{equation}
upon solving, one gets the normal vector fields for the interior and exterior spacetimes as
\begin{equation}
    n^{\mu}_{-}=\left[0,\frac{1}{a(\tau)},0,0\right],
\end{equation}
and the non-vanishing components of normalised normal vectors derived from the generalized Vaidya metric are,
\begin{equation}
    n^{v}_{+}=-\left[1-\frac{2M(r_{v},v)}{r_{v}}+2\frac{dr_{v}}{dv}\right]^{-\frac{1}{2}}
\end{equation}
and,
\begin{equation}
    n^{r_{v}}_{+}=-n^{v}_{+}\left[1-\frac{2M(r_{v},v)}{r_{v}}+\frac{dr_{v}}{dv}\right].
\end{equation}
The extrinsic curvature of the hypersurface $\Sigma$ is defined as the Lie derivative of the metric tensor with respect to the normal vector $n$, given as 
\begin{equation}
    K_{ab}=\frac{1}{2}\mathcal{L}_{n}g_{ab}=\frac{1}{2}\left[n^{c}\partial_{c} g_{ab}+ g_{cb}\partial_{a} n^{c}+g_{ac}\partial_{b} n^{c}\right].
\end{equation}
%\textcolor{red}{this is not a standard definition. The standard form is: the extrinsic curvature of the hypersurface $\Sigma$ is defined as
%\begin{equation}
%    K_{ab} = h_a^{\ c} h_b^{\ d} \nabla_c n_d,
%\end{equation}
%where $h_{ab}=g_{ab}-n_a n_b$ is the induced metric on $\Sigma$. Please check this and confirm.} \textcolor{magenta}{(AYUSH: I agree with the standard form you present. In our case, we can safely play in ADM formalism, where $h_{ij}$ is not a four-metric, unlike $g_{ab}$, but is a three-metric defined on that $\Sigma$ parametrised in time. I think one can take a Lie derivative along the metric tensor field, $g_{\mu\nu}$, (w.r.t. normal vectors $n$ defined above) without any problem. One can find a similar ADM approach in Sir Joshi's paper \cite{goswami2004nakedsingularityformationscalar} (see Eq. 35 in the paper))}.

The non-zero $\theta\theta-$component of the extrinsic curvature is given as
\begin{equation}
    K^{-}_{\theta\theta}=r_{b}a(\tau);\quad K^{+}_{\theta\theta}=r_{v}\frac{1-\frac{2M(r_{v},v)}{r_{v}}+\frac{dr_{v}}{dv}}{\left(1-\frac{2M(r_{v},v)}{r_{v}}+2\frac{dr_{v}}{dv}\right)^{\frac{1}{2}}}.
\end{equation}
Now, setting $\left[K^{-}_{\theta\theta}-K^{+}_{\theta\theta}\right]_{\Sigma}=0$ on the hypersurface $\Sigma$ together with Eq. (\ref{FieldEqn3}) and first fundamental form one gets the following relation between the mass function and Vaidya mass on the boundary as 
\begin{equation}\label{con}
    \mathcal{M}(\tau,r_{b})=2M(r_{v},v).
\end{equation}
Using the above equation, we can rewrite the condition coming from the matching of the first fundamental form as:
\begin{equation}
    \left(\frac{dv}{d\tau}\right)_{\Sigma}=\frac{1-r_{v}\dot{a}}{1-\frac{\mathcal{M}(\tau,r_{b})}{r_{v}}}.
\end{equation}
%\textcolor{red}{please check this eqn. The factor of 2 is suspicious. Normally, the Misner–Sharp mass $\mathcal{M}$ at the boundary should equal the 
%Vaidya mass $M$, without the extra factor of $2$. Thus, the correct 
%relation is
%\begin{equation}
%    \mathcal{M}(\tau,r_b) = M(r_v,v).
%\end{equation}
%I think this is either a typo or comes from a different normalization convention.}\textcolor{magenta}{I rechecked the calculations by setting $\left[K^{-}_{\theta\theta}-K^{+}_{\theta\theta}\right]_{\Sigma}=0$ and using Eq. (\ref{FieldEqn3}) and Eq. (\ref{1st}) together. I find the factor of $2$ still appears. Infact, a similar result appears in Sir Joshi's paper \cite{goswami2004nakedsingularityformationscalar} (see Eq. 37)}.

In order to get a relation describing the rate of change of Vaidya mass with respect to $r_{v}$, one has to match the $\tau$ component of the extrinsic curvature on the hypersurface $\Sigma$. Setting $\left[K^{-}_{\tau\tau}-K^{+}_{\tau\tau}\right]_{\Sigma}=0$, we get
\begin{equation}
    \frac{dM(r_{v},v)}{dr_{v}}=\frac{\mathcal{M}}{2r_{v}}+r^{2}_{b}a\ddot{a}.
\end{equation}
Now it can be seen that at the singular time the ratio $2M(r_{v},v)/r_{v}$ tends to zero. Thus, the exterior spacetime at the singular epoch reads
\begin{equation}
    ds^{2}=-dv^{2}-2dvdr_{v}+r^{2}_{v}d\Omega^{2}.
\end{equation}
The above-written metric describes a Minkowski spacetime in retarded null coordinates. Hence, the exterior generalised Vaidya metric at the singular time can be smoothly extended to the Minkowski spacetime as the collapse completes \cite{Joshi_2007}. 

Since both the first and second fundamental forms of the interior and exterior spacetimes match across the timelike hypersurface $\Sigma$, the normal derivative of the Brans–Dicke scalar field is continuous across $\Sigma$, implying $\langle \partial_n \Phi \rangle = 0$ throughout the evolution. Together with the continuity of the scalar field itself, $\langle \Phi \rangle = 0$, these conditions provide the necessary boundary conditions on $\Sigma$ for solving the Klein–Gordon equation governing $\Phi$ in the exterior generalized Vaidya spacetime.

An important implication of this Brans–Dicke collapse model is that, if the Universe is indeed governed by scalar–tensor gravity of the Brans–Dicke type, then gravitational collapse driven purely by the scalar degree of freedom is possible even for very large values of the Brans–Dicke coupling parameter $\omega$, including values $\omega \gtrsim 10^{6}$ that are fully consistent with current observational bounds.
Although large $\omega$ is often taken to imply a near–general–relativistic regime, the present model demonstrates that the dynamical evolution of the Brans–Dicke scalar field can still generate strong curvature effects in highly nonlinear collapse scenarios. In particular, the scalar field can concentrate energy density in the central region without being accompanied by a corresponding accumulation of ordinary matter, $T^{(\mathrm m)}_{ab}=0$. As a result, the collapsing configuration develops a highly energetic core purely due to scalar–field–induced curvature, ultimately terminating in a naked singularity.
As illustrated in Fig.~(\ref{BDnaked}), a finite interior region sourced solely by the Brans–Dicke scalar field ($T^{(\mathrm m)}_{ab}=0$) collapses and is smoothly matched at $r=r_b$ to an exterior generalized Vaidya spacetime. The evolution remains regular across the timelike boundary, with no surface layer, and outgoing null geodesics escape from arbitrarily close to the central singularity. The collapse therefore terminates in a naked singularity, with the exterior geometry encoding the effective radiative influence of the scalar field.

From a physical perspective, this suggests a striking observational possibility: the existence of compact, extremely energetic central regions that cannot be accounted for by the visible or baryonic matter content alone. Such configurations would mimic the gravitational influence of massive compact objects while lacking the expected matter distribution typically associated with black hole formation in general relativity. The generalized Vaidya exterior, interpreted as an effective radiating spacetime, further indicates that scalar-field dynamics may leave observable imprints through outgoing radiation or transient luminosity enhancements, even in the absence of ordinary matter flux.

This model therefore reveals a qualitatively new mechanism within scalar–tensor gravity for generating strong-field phenomena—potentially distinguishable from standard gravitational collapse in general relativity—where spacetime curvature itself, mediated by the Brans–Dicke scalar field, acts as the primary driver of collapse and singularity formation.

In the next section, we analyse the strength of the resulting singularity for both choices of the energy density, namely $\rho = 1/a$ and $\rho = -\ln a$. 

\section{Strength of the Singularity}
\label{VI}
\begin{figure*}
    \centering
    \begin{subfigure}[b]{0.48\linewidth}
        \centering
        \includegraphics[width=\linewidth]{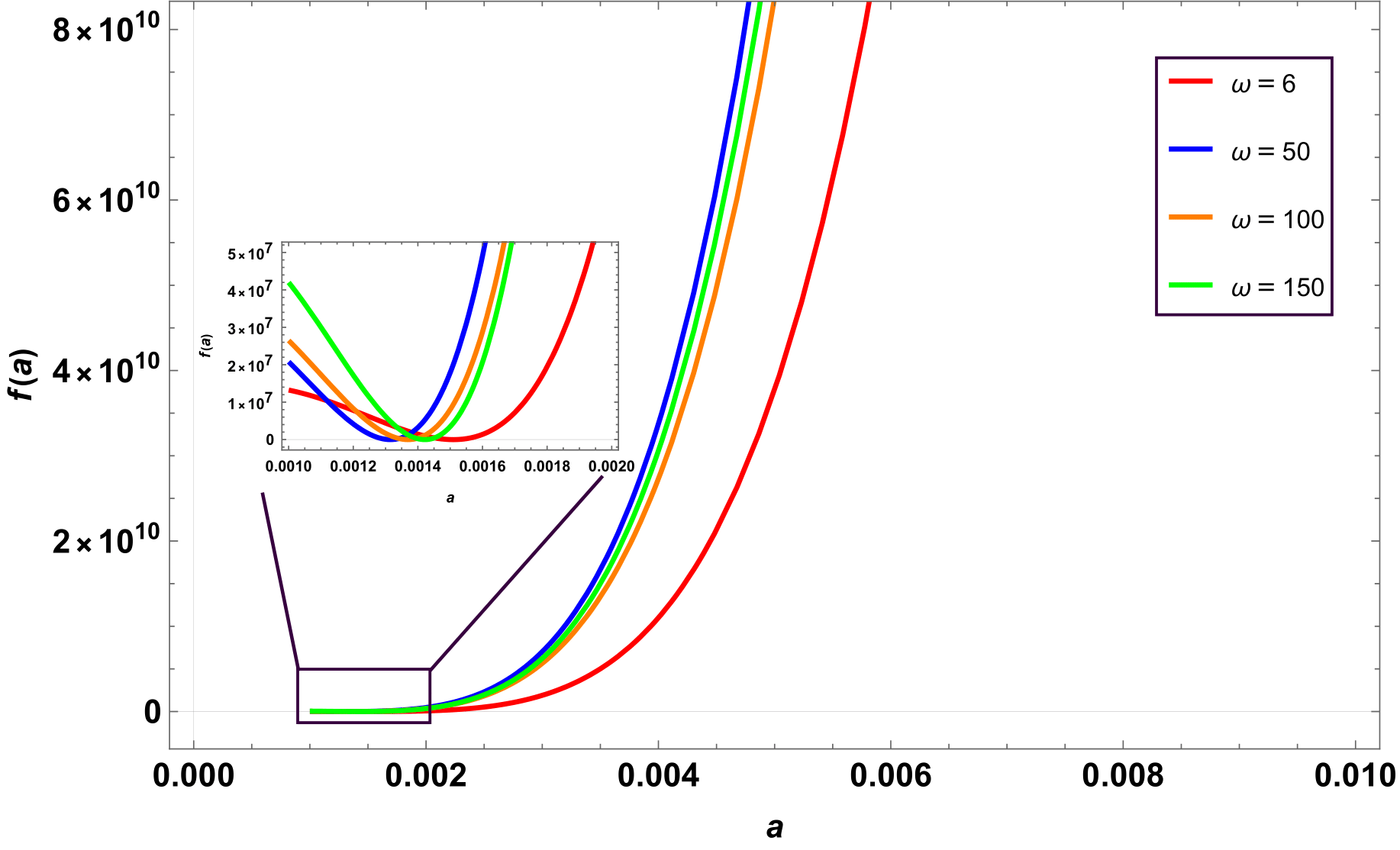}
        \caption{Plot for the LHS of the inequality in Eq. (\ref{ineq.1}) when $\rho\approx1/a$ for several values of $\omega$ such as $\omega=6$ (red), $\omega=50$ (blue), $\omega=100$ (orange), and $\omega=150$ (green), which satisfies the inequality for Tipler's strong singularity.}
        \label{Figure 4a}
    \end{subfigure}
    \begin{subfigure}[b]{0.48\linewidth}
        \centering
        \includegraphics[width=\linewidth]{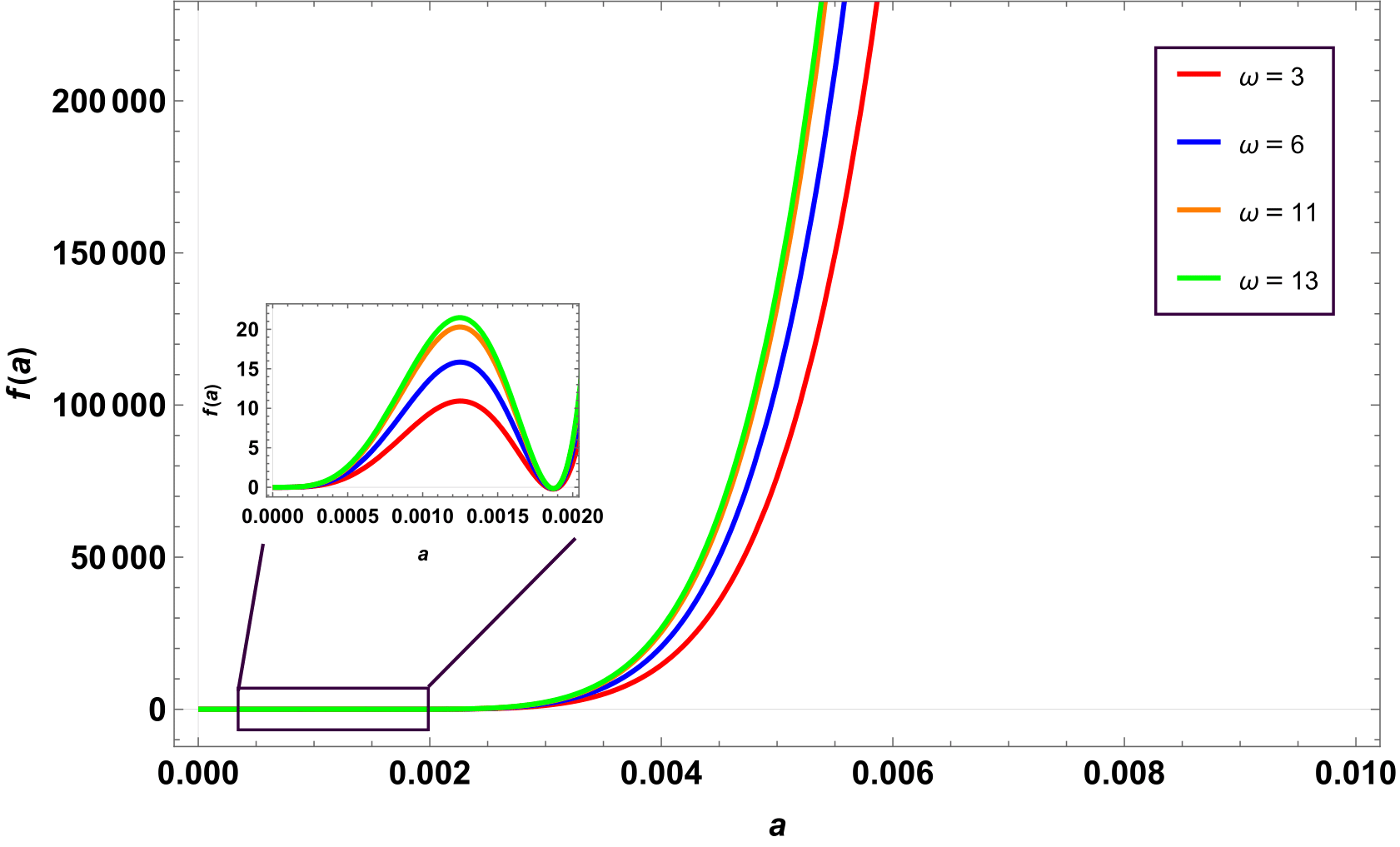}
        \caption{Plot for the inequality in Eq. (\ref{ineq.2}) when $\rho\approx-\ln{(a)}$ for several values of $\omega$ such as $\omega=3$ (red), $\omega=6$ (blue), $\omega=11$ (orange) and $\omega=13$ (green), which does not satisfies the inequality for Tipler's strong singularity.}
        \label{Figure 4b}
    \end{subfigure}
    \caption{Plots for the LHS of Tipler's strong criteria in terms of the BD scalar field, its derivatives and BD coupling parameter, for $\rho\approx1/a$ (left) and $\rho\approx-\ln{(a)}$ (right).}
    \label{Figure 4}
\end{figure*}

In general, a singularity in a spacetime is characterized by the existence of at least one incomplete causal geodesic. However, in the context of gravitational collapse, a stronger physical requirement is often imposed, namely that any extended object falling into the singularity is crushed to zero volume. A singularity satisfying this condition is said to be gravitationally strong, in the sense introduced by Tipler \cite{Tipler:1977zza}.
More precisely, let $(\mathcal{N}, g)$ be a smooth spacetime manifold and let $\alpha(\lambda)$ be a causal geodesic defined on an interval $[\lambda_0, 0)$, where $\lambda$ is an affine parameter and the endpoint $\lambda = 0$ corresponds to the singularity. Let $\chi_i$ denote a set of linearly independent Jacobi vector fields along $\alpha$, orthogonal to the tangent of the geodesic. The wedge product of these Jacobi fields defines a volume element $\mathcal{V} := \wedge \chi_i$. The singularity is said to be strong in the sense of Tipler if this volume element vanishes in the limit $\lambda \to 0$.

Clarke and Kr\'{o}lak \cite{CLARKE1985127} related the existence of a Tipler strong singularity with the growth rate of the curvature terms as follows: At least along one null geodesic with affine parameter $\lambda$ (such that $\lambda\rightarrow0$ as the singularity is approached), the following inequality holds:
\begin{equation}\label{Tipler}
\lim_{\lambda\rightarrow0}\lambda^{2}R_{\mu\nu}K^{\mu}K^{\nu}>0.
\end{equation}
Here, $R_{\mu\nu}$ is the Ricci curvature tensor, $K^{\mu}=dx^{\mu}/d\lambda$ are the tangent vectors to the null geodesics, and $x^{\mu}$ is the spacetime coordinate. This condition puts a lower bound on the growth of the Ricci curvature tensor in the spherical coordinate system, i.e. $x^{\mu}=(t,r,\theta,\phi)$. Using the FRW metric (\ref{FRW}) and the null geodesic condition $g_{\mu\nu}K^{\mu}K^{\nu}=0$, we get
\begin{equation}\label{Strn}
    K^{0}=\pm\hspace{1mm}aK^{1}
\end{equation}
During the collapse, the comoving coordinate $r$ of the collapsing shell will dynamically vary as $a\rightarrow0$. Using Eq. (\ref{Strn}) and Eq. (\ref{adot}), it is described by the following result
\begin{equation}
    r=r_0\mp\int\frac{1}{a^{2}}\sqrt{\frac{3}{\rho}}da~,
\end{equation}
where $r_0$ is the integration constant.

We construct the form of $K^{0}$ using the total derivative of $\mathcal R=ra(t)$ with respect to the affine parameter $\lambda$ in the following manner:
\begin{equation}\label{tderiv}
    \frac{d\mathcal R}{d\lambda}=a\frac{dr}{d\lambda}+r\frac{da}{d\lambda}.
\end{equation}

On solving Eq. (\ref{tderiv}) by implementing the condition in Eq. (\ref{Strn}), we obtain
\begin{equation}
    K^{0}=\pm\frac{1}{\left(1\pm r\dot{a}\right)}\frac{d\mathcal R}{d\lambda}.
    \label{K0}
\end{equation}
On the other hand, using Eq. (\ref{Strn}), we construct the form of $K^{1}$ in the following manner
\begin{equation}
    K^{1}=\frac{1}{a(1\pm r\dot{a})}\frac{d\mathcal R}{d\lambda}.
    \label{K1}
\end{equation}

Now inducing this form of $K^{0}$ and $K^{1}$ into Tipler's criteria for the strong singularity Eq. (\ref{Tipler}), namely,
\begin{equation}\label{geometry-tipler}
    \lim_{\lambda\rightarrow0}\lambda^{2}\left[R_{00}(K^{0})^{2} + R_{11}(K^{1})^{2}\right]>0.
\end{equation}
Now, we need to derive $\lim_{\lambda\to 0}\frac{d\mathcal R}{d\lambda}$ so that the above differential inequality becomes an algebraic inequality. 
For that, we can start with the Lagrangian of radial null geodesics to get the expression of $\lim_{\lambda\to 0}\frac{d\mathcal R}{d\lambda}$ in terms of the affine parameter $\lambda$: 
\begin{equation}
    \mathcal{L}\left(r,t, \frac{dr}{d\lambda}, \frac{dt}{d\lambda}\right)=\frac12\left[-\left(\frac{dt}{d\lambda}\right)^2 +a^2(t)\left(\frac{dr}{d\lambda}\right)^2\right]=0.
    \label{Lag1}
\end{equation}
Now, since $\frac{\partial \mathcal L}{\partial r}=0$, we have:
\begin{equation}
    \frac{d}{d\lambda}\left(\frac{\partial \mathcal L}{\partial(dr/d\lambda)}\right)=0 \implies a^2(t)\left(\frac{dr}{d\lambda}\right)=\mathcal C~,
    \label{Lag2}
\end{equation}
where $\mathcal C$ is a constant along the null-like curve $\alpha(\lambda)$. So, from Eq.~(\ref{Lag1}) and Eq.~(\ref{Lag2}), we get:
\begin{eqnarray}
  K^1=\left(\frac{dr}{d\lambda}\right)=\frac{\mathcal C}{a^2(t)},~ \text{and}~ K^0=\left(\frac{dt}{d\lambda}\right)=\pm \frac{\mathcal C}{a(t)}~.
  \label{K0K1}
\end{eqnarray}
Now, in the case of $\rho=\frac1a$, solving Eq.~(\ref{adot}) for $a(t)$, we can write:
\begin{eqnarray}
    a(t)=\frac{1}{12}(t_s -t)^2~,
\end{eqnarray}
where $t_s$ is the comoving time of the singularity formation, i.e., when $a(t)=0$. Now, from the expression of $\left(\frac{dt}{d\lambda}\right)$ in Eq.~(\ref{K0K1}), we can write:
\begin{eqnarray}
    \frac{d(t_s-t)}{d\lambda}=\frac{12~\mathcal C}{(t_s-t)^2}\implies \lambda =\frac{(t_s-t)^3}{36~\mathcal C}~,
    \label{lt}
\end{eqnarray}
Now, using Eq.~(\ref{tderiv}), we can get the expression of $\frac{d\mathcal R}{d\lambda}$ in terms of $(t_s - t)$ and consequently in terms of affine parameter $\lambda$:
\begin{eqnarray}
 \frac{d\mathcal R}{d\lambda} \sim -\frac{12~\mathcal C}{(t_s -t)^2}=-\frac13~(36~\mathcal C)^\frac13~\lambda^{-\frac23}~,   
\end{eqnarray}
where we use an approximation in the expression of $\frac{d\mathcal R}{d\lambda}$ as $a\to 0$ near the singularity. We can see from Eq.~(\ref{lt}) that as $t$ approaches to singular time $t_s$, $\lambda$ approaches to zero. Therefore, in the limit of the singularity (i.e., $\lambda\to 0$), $\frac{d\mathcal R}{d\lambda}\to -\infty$. Since $\lim_{\lambda\to 0}\frac{d\mathcal R}{d\lambda}= -\infty$, using L'Hôpital rule (i.e., $\lim_{\lambda\to 0}\frac{\mathcal R}{\lambda}=\lim_{\lambda\to 0}\frac{d\mathcal R}{d\lambda}$), one cannot conclude $\lim_{\lambda\to 0}\left(\frac{\mathcal R}{\lambda}/\frac{d\mathcal R}{d\lambda}\right)=1$.  One has to separately derive the expression of $\frac{\mathcal{R}}{\lambda}$ near the singularity:
\begin{eqnarray}
    \frac{\mathcal R}{\lambda}\sim -(36~\mathcal C)^\frac13~\lambda^{-\frac23}\, ,
\end{eqnarray}
which implies near the singularity:
\begin{eqnarray}
    \frac{d\mathcal R}{d\lambda}=\frac13 \frac{\mathcal R}{\lambda}~.
\end{eqnarray}
Now, we will use the above expression of $\frac{d\mathcal R}{d\lambda}$ along with the expressions of $K^0$ and $K^1$ in Eq.~(\ref{K0}) and Eq.~(\ref{K1}), respectively to get the expression of $\lim_{\lambda\rightarrow0}\lambda^{2}\left[R_{00}(K^{0})^{2} + R_{11}(K^{1})^{2}\right]$:
\begin{eqnarray}
    \lim_{\lambda\rightarrow0}\lambda^{2}\left[R_{00}(K^{0})^{2} + R_{11}(K^{1})^{2}\right]&=&\lim_{a\to 0}\frac{2~r^2}{9(1\pm r\dot{a})^2}\left(\dot{a}^{2}-\ddot{a}a\right)\nonumber\\&=&\frac49\, ,
\end{eqnarray}
where we have used the following expressions of the 00-component and 11-component of the Ricci tensor for the FRW metric (\ref{FRW}):
\begin{equation}
R_{00}=-3\frac{\ddot{a}}{a};\quad R_{11}=(a\ddot{a}+2\dot{a}^{2})\, .
\end{equation}
On the other hand, For $\rho\approx-\ln{a}$, the expression of scale factor using $\dot{a}=-a\sqrt{(-\ln{a})/3}$ can be written as:
\begin{equation*}
 a(t)=\exp{\left(-\frac{1}{12}(t_{s}-t)^{2}\right)}.
\end{equation*}
It can be seen that as $t\rightarrow t_{s}$, the scale factors tend to a positive value, i.e., $a(t)\rightarrow1$. Expressing the $K^{0}$ vector as using the scale factor $a(t)$ for $\rho\approx-\ln{(a)}$, we get:
\begin{equation}
    \frac{d(t_{s}-t)}{d\lambda}=\frac{\mathcal{C}}{\exp{\left(-(t_{s}-t)^{2}/12\right)}}.
\end{equation}
Integrating both sides provides us with the expression for the affine parameter $\lambda$ as a function of $(t_{s}-t)$, namely:
\begin{equation}
    \lambda=\frac{\sqrt{3\pi}}{\mathcal{C}}\hspace{1mm}\text{Erf}\left(\frac{t_{s}-t}{2\sqrt{3}}\right).
\end{equation}
This can also be written as:
\begin{equation}
    (t_{s}-t)=2\sqrt{3}\hspace{1mm}\text{Erf}^{-1}\left(\frac{\mathcal{C}\lambda}{\sqrt{3\pi}}\right).
\end{equation}
We can now write the scale factor in terms of $\lambda$ as:
\begin{equation}
    a(\lambda)=\exp\Bigg\{-\left[\text{Erf}^{-1}\left(\frac{\mathcal{C}\lambda}{\sqrt{3\pi}}\right)\right]^{2}\Bigg\}.
\end{equation}
On the other hand, the radial comoving coordinates $r$ in terms of $\lambda$ becomes:
\begin{equation}
\begin{aligned}
     r(\lambda) = \sqrt{3\pi}\hspace{1mm}\text{Erfi}\Bigg\{\text{Erf}^{-1}\left(\frac{\mathcal{C}\lambda}{\sqrt{3\pi}}\right)\Bigg\}.
\end{aligned}
\end{equation}
For our scenario $\lambda\rightarrow0$\footnote{The Taylor expansion of $\text{Erf}^{-1}(x)$ for very small $x$ can be written as:
\begin{equation*}
    \text{Erf}^{-1}(x)\approx \frac{\sqrt{\pi}}{2}\left(x+\frac{\pi x^{3}}{12} +\dots\right)
\end{equation*}}, the expressions for $a(\lambda)$ and $r(\lambda)$ takes the following form:
\begin{equation}
    a(\lambda)=\exp\left(-\frac{\mathcal{C}^{2}\lambda^{2}}{12}\right).
\end{equation}
and
\begin{equation}
    r(\lambda)=\sqrt{3\pi}\hspace{1mm}\text{Erfi}\left(\frac{\mathcal{C}\lambda}{2\sqrt{3}}\right).
\end{equation}
We also take the asymptotic limit of the $\text{Erfi}(x)$ function\footnote{The asymptotic series expansion of the $\text{Erfi}(x)$ function up to first order can be written as:\begin{equation*}
    \text{Erfi}(x)\approx \frac{e^{-x^{2}}}{x\sqrt{\pi}}
\end{equation*}} and find the final expression of $r(\lambda)$ as:
\begin{equation}
    r(\lambda)=\frac{6}{\mathcal{C}\lambda}\exp{\left(-\frac{\mathcal{C}^{2}\lambda^{2}}{12}\right)}.
\end{equation}
Now we write down the expanded form of $d\mathcal{R}/d\lambda$ as:
\begin{equation}
    \begin{aligned}
        \frac{d\mathcal{R}}{d\lambda}&=-\frac{2\left(\mathcal{C}^2 \lambda ^2+3\right)}{\mathcal{C} \lambda ^2}\exp{\left(-\frac{1}{6}\mathcal{C}^2 \lambda ^2\right)}.
    \end{aligned}
\end{equation}
In addition, writing the expression for $\mathcal{R}/\lambda$ as:
\begin{equation}
    \begin{aligned}
        \frac{\mathcal{R}}{\lambda}&=\frac{6}{\mathcal{C}\lambda ^2}\exp{\left(-\frac{1}{6} \mathcal{C}^2 \lambda ^2\right)}.
    \end{aligned}
\end{equation}
We get:
\begin{equation}\label{drdlambda}
    \frac{d\mathcal{R}}{d\lambda}=-\left(\frac{\mathcal{C}^2 \lambda ^2}{3}+1\right)\frac{\mathcal{R}}{\lambda}.
\end{equation}
Within the limit $\lambda\rightarrow0$, Eq. (\ref{drdlambda}) becomes,
\begin{equation}
    \lim_{\lambda\rightarrow0}\frac{d\mathcal{R}}{d\lambda}=-\frac{\mathcal{R}}{\lambda}.
\end{equation}
 
Now, for $\rho=-\ln{a}$, using $\dot{a}=-a\sqrt{-\ln{a}/3}$, $r=\sqrt{3\pi}\hspace{1mm}\text{Erfi}(\sqrt{-\ln{a}})$, $r\dot{a}=-a\sqrt{-\pi\ln{a}}\hspace{1mm}\text{Erfi}(\sqrt{-\ln{a}})$, $\ddot{a}=-\left[(a\ln{a})/3+a/6\right]$, we get: 
    \begin{equation}
    \lim_{a\rightarrow0^{+}}\frac{a^{2}\pi\hspace{1mm}[\text{Erfi}(\sqrt{-\ln{a}})]^{2}}{\left(1-a\sqrt{\pi(-\ln{a})}\hspace{1mm}\text{Erfi}(\sqrt{-\ln{a}})\right)^{2}}\rightarrow 0.
    \end{equation}

Therefore, this does not satisfy Tipler's criteria, indicating that the singularity formed for $\rho\approx-\ln{(a)}$ is \textit{weak}. We retrieve that Tipler's strong singularity criteria are satisfied only when $\rho\approx1/a$, while for $\rho\approx-\ln{(a)}$ the strength of the singularity is weak. Furthermore, this geometrical approach can also be extended towards constructing a Tipler's strong singularity criteria in terms of the BD scalar field $\Phi(a)$, and it can be achieved by first estimating the Ricci scalar, namely $R=g^{\mu\nu}R_{\mu\nu}$ and then substituting it back into the field equation (Eq. (\ref{FieldEqn1})). The Ricci scalar can be written as
\begin{equation}
    \begin{split}
        R=-\frac{T^{\text{m}}}{\Phi}-\frac{\omega}{\Phi^{2}}\left(g^{\mu\nu}\nabla_{\mu}\Phi\nabla_{\nu}\Phi-2\nabla_{\eta}\Phi\nabla^{\eta}\Phi\right)-\\\frac{1}{\Phi}\left(g^{\mu\nu}\nabla_{\mu}\nabla_{\nu}\Phi-4\Box\Phi\right)+2\frac{V(\Phi)}{\Phi}.
        \end{split}
\end{equation}

Here, $T^{\text{m}}=g^{\mu\nu}T^{(\text{m})}_{\mu\nu}$, and also taking into account the assumption $T^{\text{m}}_{\hspace{1mm}\mu\nu}\rightarrow 0$ (absence of matter). Now putting this back into Eq. (\ref{FieldEqn1}). One obtains the following expression for the Ricci tensor
\begin{equation}
\begin{split}
    R_{\alpha\beta}=\frac{\omega}{\Phi^{2}}\left(\nabla_{\alpha}\Phi\nabla_{\beta}\Phi-\frac{1}{2}g_{\alpha\beta}\nabla_{\eta}\Phi\nabla^{\eta}\Phi\right)\\-\frac{g_{\alpha\beta}}{2}\frac{\omega}{\Phi^{2}}\left(g^{\mu\nu}\nabla_{\mu}\Phi\nabla_{\nu}\Phi-2\nabla_{\eta}\Phi\nabla^{\eta}\Phi\right)\\-\frac{g_{\alpha\beta}}{2\Phi}\left(g^{\mu\nu}\nabla_{\mu}\nabla_{\nu}\Phi-4\Box
    \Phi\right)\\+\frac{1}{\Phi}\left(\nabla_{\alpha}\nabla_{\beta}\Phi-g_{\alpha\beta}\Box\Phi\right).
\end{split}    
\end{equation}
Let us now substitute this expression for the Ricci tensor in the inequality (Eq. (\ref{Tipler})). We then obtain
\begin{equation}
    \begin{split}
        \lim_{\lambda\rightarrow0}\lambda^{2}\left[\frac{\omega}{\Phi^{2}}\left(\nabla_{\alpha}\Phi\nabla_{\beta}\Phi-\frac{1}{2}g_{\alpha\beta}\nabla_{\eta}\Phi\nabla^{\eta}\Phi\right)\right]K^{\alpha}K^{\beta}\\+\lim_{\lambda\rightarrow0}\lambda^{2}\left[\frac{g_{\alpha\beta}}{2}\frac{\omega}{\Phi^{2}}\left(2\nabla_{\eta}\Phi\nabla^{\eta}\Phi-g^{\mu\nu}\nabla_{\mu}\Phi\nabla_{\nu}\Phi\right)\right]K^{\alpha}K^{\beta}\\+\lim_{\lambda\rightarrow0}\lambda^{2}\left[\frac{g_{\alpha\beta}}{2\Phi}\left(4\Box
    \Phi-g^{\mu\nu}\nabla_{\mu}\nabla_{\nu}\Phi\right)\right]K^{\alpha}K^{\beta}\\+\lim_{\lambda\rightarrow0}\lambda^{2}\left[\frac{1}{\Phi}\left(\nabla_{\alpha}\nabla_{\beta}\Phi-g_{\alpha\beta}\Box\Phi\right)\right]K^{\alpha}K^{\beta}>0.
    \end{split}
\end{equation}
For null geodesics $g_{\alpha\beta}K^{\alpha}K^{\beta}=0$. This simplifies our inequality in the following manner
\begin{equation}\label{Tipler1}
\begin{split}\lim_{\lambda\rightarrow0}\lambda^{2}\left[\frac{\omega}{\Phi^{2}}\left(\nabla_{\alpha}\Phi\nabla_{\beta}\Phi\right)+\frac{1}{\Phi}\left(\nabla_{\alpha}\nabla_{\beta}\Phi\right)\right]K^{\alpha}K^{\beta}>0.
    \end{split}
\end{equation}
Now we look to incorporate our scenario where we impose the assumptions on the expression for $\rho$. We will first derive the general results in the form of $\rho$ and then insert their specific expressions. Using the derived expressions for $K^{0}$ and $K^{1}$, the generalised expression for Tipler's strong singularity turns out to be as follows:
\begin{equation}
\lim_{\lambda\rightarrow0}\frac{\lambda^{2}}{(1+r\dot{a})^{2}}\left[\frac{\omega}{\Phi^{2}}\dot{\Phi}^{2}+\frac{\ddot{\Phi}}{\Phi}\right]\left(\frac{d\mathcal{R}}{d\lambda}\right)^{2}>0.
\end{equation}
Introducing the $\dot{a}$ and $\ddot{a}$ expressions into this inequality gives the following result:
\begin{equation}
   \lim_{\lambda\rightarrow0}\frac{\lambda^{2}}{(1+r\dot{a})^{2}}\left[\frac{\omega\dot{a}^{2}\Phi^{2}_{,a}}{\Phi^{2}}+\frac{\ddot{a}\Phi_{,a}}{\Phi}+\frac{\dot{a}^{2}\Phi_{,aa}}{\Phi}\right]\left(\frac{d\mathcal{R}}{d\lambda}\right)^{2}>0.
\end{equation}
The above-written inequality form of the Tipler's strong singularity criteria is written in terms of the BD scalar field $\Phi(a)$. Moving forward, we now plug in the expressions for $\rho\approx1/a$ and then $\rho\approx-\ln{(a)}$ as:
\begin{itemize}
    \item When $\rho\approx1/a$, the LHS of the inequality becomes
    \begin{equation}\label{ineq.1}
        \lim_{a\rightarrow0^{+}}\frac{4a}{3}\left[\frac{\omega a\Phi^{2}_{,a}}{3\Phi^{2}}+
        \frac{\Phi_{,a}}{6\Phi}+\frac{a\Phi_{,aa}}{3\Phi}\right].
    \end{equation}
    
%For $\rho\approx1/a$, we observe that the singularity appears to be strong as it approaches a positive finite value near the singularity. Moreover, within our numerical analysis we observe that the singularity is strong for all $\omega>0$ parameter space. We depict the results in the Figure (\ref{Figure 4a}).

For $\rho \approx 1/a$, we find that the singularity is strong, with the quantity $\frac{4a}{3}\left[\frac{\omega a\Phi^{2}_{,a}}{3\Phi^{2}}+
        \frac{\Phi_{,a}}{6\Phi}+\frac{a\Phi_{,aa}}{3\Phi}\right]$ approaching a positive finite value as the singularity is approached. Moreover, our numerical analysis indicates that the singularity remains strong throughout the entire parameter range $\omega > 0$. These results are illustrated in Fig.~(\ref{Figure 4a}).
    \item When $\rho\approx-\ln{(a)}$, the LHS of the inequality becomes
    \begin{equation}\label{ineq.2}
    \begin{split}
        \lim_{a\rightarrow0^{+}}- \frac{3\pi a^{2}\left[\text{Erfi}(\sqrt{-\ln{a}})\right]^{2}}{
        \left(1-a\sqrt{\pi(-\ln{a})}\text{Erfi}(\sqrt{-\ln{a}})\right)^{2}
        } \Bigg[
            \frac{\omega a^{2}\ln{a}\Phi^{2}_{,a}}{3\Phi^{2}}\\
            + \left(\frac{a\ln{a}}{3}+\frac{a}{6}\right)\frac{\Phi_{,a}}{\Phi} 
            + \frac{a^{2}\ln{a}\Phi_{,aa}}{3\Phi}
        \Bigg].
    \end{split}
\end{equation}
For $\rho\approx-\ln{(a)}$, we observe that the singularity is weak, as the LHS of Tipler's inequality approaches zero near the singularity, only for specific values of $\omega$, such as $\omega=1,\dots,7,11,12,13,15,\dots,18,24,27,\dots,32,39,49$ (see Figure \ref{Figure 4b}).
\end{itemize}

\section{Conclusions}
\label{VII}
In this work, we have systematically studied gravitational collapse within the framework of BD gravity in the absence of conventional matter fields. Specifically, we considered a scenario in which the collapse is driven solely by the dynamical evolution of the BD scalar field $\Phi(t)$. Such a collapse is a characteristic feature of extended theories of gravity beyond GR, which introduce additional degrees of freedom. In the present case, the extra dynamical variable is the BD scalar field, or equivalently the time-dependent effective gravitational coupling $G(t) \simeq \Phi^{-1}(t)$. Our results highlight the unique role of scalar-tensor modifications of gravity in producing novel outcomes for the nature and visibility of spacetime singularities.

Following the approaches established in \cite{goswami2004nakedsingularityformationscalar, Goswami:2005fu, Goswami:2007na, mosani}, we constructed our collapse model through employment of the energy density expression $\rho\approx1/a$ and $\rho\approx-\ln{(a)}$, which become unbounded and diverge to infinity near the singularity $a\rightarrow0$. In particular, we restrict our analysis to the region $ a\ll 1$. Our numerical results as illustrated in Figures (\ref{Figure 1}) and (\ref{Fig2}) for $\rho\approx1/a$ and Figures (\ref{Fig3}) and (\ref{Fig4}) for $\rho\approx-\ln{(a)}$, highlight the behaviour of the BD field $\Phi$ and $V(\Phi)$ near the singularity. Let us reiterate that we analysed for $\rho\approx1/a$ with coupling parameter $\omega=6$ and $\omega=10^{6}$. Notably, the latter value is chosen according to the observational constraint on $\omega$. We showed that the BD scalar field $\Phi(a)\rightarrow+\infty$ as the scale factor $a\rightarrow0$ near the singularity (see Figure (\ref{Figure 1})). In addition, we presented the evolution of the self-interaction potential $V(\Phi)$ as $\Phi\rightarrow+\infty$, and observed that $V(\Phi)\rightarrow+\infty$ near the singularity (see Figure (\ref{Fig2})). On the other hand, when $\rho\approx-\ln{(a)}$ for $\omega=6$ and $\omega=10^{6}$, we outlined the plot of the BD scalar field $\Phi(a)\rightarrow+\infty$ near the singularity quite similarly to how $\Phi(a)$ diverges when $\rho\approx1/a$ (see Figure (\ref{Fig3})). However, the evolution of $V(\Phi)$ in the case asymptotically approaches some finite negative value near the singular region (see Figure (\ref{Fig4})). Moreover, we also provide the theoretical expressions for the Kretschmann invariant as a function of the scale factor and highlight its diverging nature, indicating the presence of a curvature singularity for both cases, i.e., $\rho\approx1/a$ and $\rho\approx-\ln{(a)}$. In brief, this analysis confirmed that the BD scalar field alone is sufficient to drive the system to a singular state, without the need for any additional matter contributions.

We then examined the causal structure of the singularity to determine whether it is visible or hidden. Specifically, we investigated the existence of outgoing causal curves originating from
the singularity. Our analysis demonstrated that for $\rho\approx1/a$ and $\rho\approx-\ln{(a)}$, a globally visible naked singularity is formed for $r_{b}<\sqrt{3}$ and $r_{b}<\frac{\sqrt{3}}{\text{max}\left(a\sqrt{|\ln{a}|}\right)}\sim4.0385$, respectively, allowing future-directed nonspacelike geodesics emerging from it to reach the asymptotic null infinity. While for $r_{b}>3$ and $r_{b}>4.0385$ we obtain $\mathcal{M}\geq\mathcal{R}$ for finite time interval within $[r_{MTS}(t),r_{b}]$. As the collapse proceeds, $r_{MTS}(t)$ increases till it reaches $r_{b}$, making the spacetime region free of trapped surfaces, allowing the singularity to be naked. We emphasise that such a disappearance of the trapped band surfaces can lead to a sharp change in the internal dynamical state, allowing a sudden outflux of null rays propagating asymptotically towards null infinity (see Figure (\ref{BDnaked}) for a pictorial representation). Thus, the collapse leads to the formation of a naked singularity rather than a black hole.

To further quantify the nature of the singularity, within our collapsing model, we constructed the Israel junction conditions by smoothly matching the interior flat ($k=0$) FRW spacetime (\ref{interior}) with the generalised Vaidya exterior spacetime for outgoing null radiation (\ref{exterior}). We then proceeded to analyse the strength of the naked singularity formed for both energy density profiles, namely $\rho \approx 1/a$ and $\rho \approx -\ln(a)$. Notably, within our collapsing model, we worked with newly constructed mathematical expressions for $K^{0}$, $K^{1}$, and $\lim_{\lambda\rightarrow0}d\mathcal{R}/d\lambda$, which differ from the ones constructed in \cite{mosani} for the related investigation. In particular, we cleanly derive the above-mentioned quantities before analysing the strength of the singularity. And then, only we proceed to address a ``two-fold" query. Firstly, we particularly begin by analysing the geometric construction of Tipler's inequality (\ref{Tipler}) and conclude that the naked singularity formed for $\rho\approx 1/a$ is Tipler strong, while the naked singularity is \textit{not} Tipler strong for $\rho\approx-\ln{(a)}$. Secondly, we extend Tipler's inequality in terms of the BD scalar field $\Phi(a)$, its derivatives and the BD coupling parameter $\omega$ for both the energy density profiles, corroborated with the numerical plots showcasing the behaviour of the LHS of Tipler's inequality near the singularity in Figure (\ref{Figure 4}). Based on our analysis, we concluded that for $\rho\approx 1/a$ the LHS of Tipler's inequality asymptotes towards a positive finite value satisfying Tipler's strong criteria for all positive values of $\omega$. On the other hand, for $\rho\approx-\ln{(a)}$, the LHS of Tipler's inequality asymptotes towards zero/non-positive values for specific values of $\omega$, indicating a violation of Tipler's strong criteria, i.e., the singularity is \textit{weak}. In conclusion, we find a promising consistency between our analytic and numerical results throughout our geometric and BD scalar field analysis, based on the strength of the singularity in Tipler's sense.

The formation of a globally naked singularity in our model reveals deeper insights into the physical and observational implications. 
Specifically, the spacetime curvature is mediated purely through the dynamics of the Brans-Dicke scalar field, which serves as the fundamental agent driving both the collapse process and the ultimate formation of a naked singularity. 
As depicted in Figure (\ref{BDnaked}), the collapse scenario proceeds through three distinct evolutionary stages. Initially, a finite interior region characterized by the metric $g^-_{\mu\nu}$ and sourced exclusively by the Brans-Dicke scalar field ($T^{(\mathrm{m})}_{ab}=0$) is enclosed inside the spherical timelike surface of radius $r=r_b$, beyond which the exterior spacetime is described by the metric $g^{\mathrm{+}}_{\mu\nu}$ with ordinary matter contributions ($T^{(\mathrm{m})}_{\mu\nu}\neq 0$). As the collapse progresses, the interior domain contracts while maintaining continuous junction conditions at the timelike boundary $r=r_b$, which smoothly joins the scalar-field-dominated interior $g^-_{\mu\nu}$ to the exterior generalised Vaidya spacetime $g^+_{\mu\nu}$. In the final stage, the interior region has completely collapsed, forming a central naked singularity from which outgoing null geodesics (indicated by the red trajectories) successfully escape to the exterior, confirming the absence of an event horizon. Throughout this evolution, the dynamics remain regular across the matching hypersurface with no surface layer discontinuities, while the exterior geometry encodes the effective radiative influence of the scalar field dynamics. The escaping null geodesics ensure that quantum information from the singularity remains accessible, through continuous radiative encoding in the generalised Vaidya exterior. Moreover, this violates the cosmic censorship conjecture (CCC) by demonstrating physically reasonable initial data (smooth scalar field profiles) that evolve into naked singularities, providing a concrete counterexample where strong curvature regions remain causally naked rather than hidden behind horizons. This scenario reveals intriguing astrophysical implications: the formation of such ultra-compact objects includes high-energy cores whose extreme gravitational effects cannot be explained by baryonic or electromagnetically visible matter distributions alone. These structures would replicate the lensing and dynamical signatures of massive compact objects while exhibiting fundamentally different matter profiles than those expected from standard general relativistic collapse to black holes. Building upon these results, in future works, one can explore the role of deformations in the configurational phase space of the theory, investigating whether naked singularities persist under such modifications (see \cite{Rasouli_2014} for a similar treatment in GR, extending the work done in \cite{goswami2004nakedsingularityformationscalar} through means of deformations). Another promising direction is the inclusion of loop quantum gravity (LQG) corrections within this collapse model to investigate whether quantum corrections can mitigate the occurrence of a singularity due to sudden outflux of matter, as shown in \cite{Goswami:2005fu} for GR. Furthermore, the effects of other matter forms, such as dust, cosmic strings, and radiation, can be incorporated into scalar-tensor theories to explore more general collapse dynamics beyond the purely scalar-field case. In conclusion, our study demonstrates the rich physics that BD gravity theory entails for gravitationally collapsing scenarios driven entirely by the BD scalar field, culminating in the formation of a naked singularity. These singularities are both locally and globally visible. Notably, these results underscore the rich phenomenology of scalar-tensor theories and highlight potential avenues for testing deviations from general relativity through astrophysical and cosmological observations of extreme gravitational environments.

%\textbf{\textcolor{red}{\MakeUppercase{This paragraph can be adjusted in the above paragraph since this is related to the observational implications and combined these two to make one paragraph.}}}

\section*{Acknowledgements}
A. Bidlan would like to acknowledge the contribution of the COST Action \textbf{CA23130} (Bridging high and low energies in search of quantum gravity (BridgeQG)) community. P. Bambhaniya acknowledge support from the São Paulo Research Foundation (FAPESP) under grants No. \textbf{2024/09383-4}.).
%\bibliography{apssamp}
\bibliography{apssamp_updated}
\end{document}